\documentclass{emulateapj}
\usepackage{amssymb, amsmath, epsfig}

\newcommand{\bfk}{{\boldsymbol{k}}}
\newcommand{\xvec}{{\boldsymbol{x}}}

\newcommand{\bfP}{{\boldsymbol{P}}}

\newcommand{\bff}{{\boldsymbol{f}}}
\newcommand{\bfn}{{\boldsymbol{n}}}
\newcommand{\bfs}{{\boldsymbol{s}}}
\newcommand{\bfI}{{\boldsymbol{I}}}
\newcommand{\bfmu}{\mbox{\boldmath$\mu$}}
\newcommand{\bfnu}{\mbox{\boldmath$\nu$}}
\newcommand{\bfu}{{\boldsymbol{u}}}
\newcommand{\bfr}{{\boldsymbol{r}}}

\newcommand{\bfF}{{\boldsymbol{F}}}
\newcommand{\bfC}{{\boldsymbol{C}}}

\newcommand{\bfx}{{\boldsymbol{x}}}
\newcommand{\kvec}{{\boldsymbol{k}}}
\newcommand{\uvec}{{\boldsymbol{u}}}
\newcommand{\nhat}{{\hat{\boldsymbol{n}}}}
\newcommand{\Mpc}{{\rm Mpc}}
\newcommand{\MHz}{{\rm MHz}}

\newcommand{\meter}{{\rm m}}

\begin{document}

\title{Cosmological Parameter Estimation Using 21 cm Radiation from
the Epoch of Reionization} \author{Matthew McQuinn\altaffilmark{1}, Oliver Zahn\altaffilmark{1},
Matias Zaldarriaga\altaffilmark{1,2},
Lars Hernquist\altaffilmark{1},
 Steven R. Furlanetto\altaffilmark{3}}

\altaffiltext{1} {Harvard-Smithsonian Center for Astrophysics, 60
Garden St., Cambridge, MA 02138; mmcquinn@cfa.harvard.edu}

\altaffiltext{2} {Jefferson Laboratory of Physics, Harvard University,
Cambridge, MA 02138}

\altaffiltext{3} {Division of Physics, Mathematics, \& Astronomy;
  California Institute of Technology;  Pasadena, CA
  91125}

\begin{abstract}

A number of radio interferometers are currently being planned or
constructed to observe 21 cm emission from reionization.  Not only
will such measurements provide a detailed view of that epoch, but,
since the 21 cm emission also traces the distribution of matter in the
Universe, this signal can be used to constrain cosmological parameters
at $6 \la z \la 20$.  The sensitivity of an interferometer to the
cosmological information in the signal may depend on how precisely the
angular dependence of the 21 cm 3-D power spectrum can be measured.
Utilizing an analytic model for reionization, we quantify all the
effects that break the spherical symmetry of the 3-D 21 cm power
spectrum and produce physically motivated predictions for this power
spectrum.  We find that upcoming observatories will be sensitive to
the 21 cm signal over a wide range of scales, from larger than $100$
to as small as $1$ comoving Mpc.  Next, we consider three methods to
measure cosmological parameters from the signal: (1) direct fitting of
the density power spectrum to the signal (this method can only be
applied when density fluctuations dominate the 21 cm fluctuations),
(2) using only the velocity field fluctuations in the signal, (3)
looking at the signal at large enough scales such that all
fluctuations trace the density field.  With the foremost method, the
first generation of 21 cm observations should moderately improve
existing constraints on cosmological parameters for certain
low-redshift reionization scenarios, and a two year observation with
the second generation interferometer MWA5000 in combination with the
CMB telescope Planck can improve constraints on $\Omega_w$ (to $\pm
0.017$, a $1.7$ times smaller uncertainty than from Planck alone),
$\Omega_m \, h^2$ ($\pm 0.0009$, $2.5$ times), $\Omega_b \,h^2$ ($\pm
0.00012$, $1.5$ times), $\Omega_{\nu}$ ($\pm0.003$, $3$ times), $n_s$
($\pm 0.0033$, $1.4$ times), and $\alpha_s$ ($\pm 0.003$,
$2.7$ times).  Larger interferometers, such as SKA, have the potential
to do even better.  If the Universe is substantially ionized by $z
\sim 12$ or if spin temperature fluctuations are important, we show
that it will be difficult to place competitive constraints on
cosmological parameters from the 21 cm signal with any of the
considered methods.

\end{abstract}


\section{Introduction}
\label{intro}

The reionization of the Universe involves many poorly understood
astrophysical phenomena such as the formation of stars, the escape
of ionizing photons from star-forming regions, and the
evolving clumpiness of the gas in the intergalactic medium (IGM).
However, reionization imprints signatures onto 21 cm emission from
high-redshift neutral hydrogen, as will be studied with the instruments
PAST, LOFAR, and MWA, in a manner that is sensitive to these
processes.\footnote{For more information, see http://www.lofar.org/,
http://web.haystack.mit.edu/arrays/MWA/, and \citet{pen04}.}
Moreover, the 21 cm emission encodes information pertaining to
fundamental cosmological parameters.  Due to all the overlying
astrophysics, it is uncertain whether or not 21 cm observations can be
competitive with other cosmological probes.

Several authors have discussed using the 21 cm signal from
reionization to study cosmology, in addition to mapping out the
epoch of reionization (EOR); \citep{scott90, tozzi00, iliev02}.
Recently, \citet{barkana04a} show that redshift-space distortions
from peculiar velocities allow for the decomposition of the
observed 21 cm 3-D power spectrum into terms that are proportional
to $\mu^0, \mu^2$ and $\mu^4$, where $\mu$ is the cosine of the
angle between a mode $\kvec$ and the line-of-sight (LOS). In
principle, this decomposition makes it possible to separate the
contribution from reionized bubbles from that owing to a fundamental
cosmological quantity, the linear-theory density power spectrum.

Even if the signal from the ionized bubbles dominates over the
cosmological one, \citet{nusser04} shows that one can look for an
asymmetry between the 21 cm signal measured in depth and that
measured in angle.  The presence of this asymmetry may imply that the
cosmology assumed in the analysis is incorrect [the Alcock-Paczynski (AP)
effect].  This effect could help further constrain $\Omega_m$ and
$h$, as well as dark energy models \citep{nusser04}.  It might be
possible to distinguish the AP effect because it creates a $\mu^6$
dependence in the 3-D power spectrum, which is distinct from the
behavior that arises from velocity-field effects alone
\citep{nusser04, barkanaAP}.

For both the $\mu$ decomposition of the 21 cm power spectrum and the
AP effect, the feasibility of inferring cosmological parameters using
future surveys depends on how sensitive these surveys are to
deviations from spherical symmetry in the 3-D power spectrum.
\citet{morales05} suggests that 21 cm observations should spherically
average $\bfk$-modes over a shell in Fourier space to increase the
signal-to-noise ratio.  In addition to losing the angular information
contained in the signal, such averaging would significantly bias
upcoming measurements: The power spectrum is not close to spherically
symmetric and 21 cm interferometers will be most sensitive to modes
oriented along the LOS.  Array design also factors into the sensitivity
to certain features in the signal.  The first generation of EOR arrays
are still being planned, and so it is important to understand different
design trade-offs.

Authors have considered the 21 cm emission signal in several limits,
such as when the spin temperature fluctuations are still important and
the HII regions are small \citep{barkana05, ahn05} or when the spin
temperature is much larger than the cosmic microwave background (CMB)
temperature \citep{zald04, furlanetto04a, santos05}.  In reality, we
do not know how quickly X-rays from the first stars, black holes and
shocks will heat the gas and how quickly the spin temperature can
couple to the gas temperature via the Wouthuysen-Field effect
\citep{wout, field58}.  Previous work suggests that these processes
act quickly after the first stars turn on \citep{oh01, venkatesan01,
chen04,ciardi03-21cm}.  Moreover, it is expected that quasars or
stellar sources contribute the bulk of the ionizing radiation.  If this
is the case, reionization will be a patchy process, in which the HII
regions grow around these sources as the ionization fraction
increases.

Upcoming observations will be most sensitive to lower redshifts ($z
\sim 6-12$) during reionization \citep{bowman05}.  At these low
redshifts, it is likely that the spin temperature is greater than the
CMB temperature and that the ionized fraction is of order unity and
very patchy.  This is the regime that we consider for much of this paper.
It is also possible that the ionization fraction is near zero for a
period at these low redshifts, which will facilitate cosmological
parameter estimation.  We consider this case as well.

The organization of this paper is as follows.  In \S \ref{21cm} we
make physically motivated predictions for the form of the 3-D 21 cm power
spectrum.  We then generalize the detector noise calculation of
\citet{morales05} to a power spectrum that is not spherically
symmetric (\S \ref{noise}) and incorporate realistic foregrounds into this
sensitivity calculation (\S \ref{foregrounds}).  These calculations
allow us to estimate the sensitivity of upcoming interferometers to
the 21 cm power spectrum (\S \ref{detectors}).  We conclude with a discussion of how
the 21 cm signal can be used to measure fundamental cosmological parameters as
well as a Fisher matrix analysis to estimate how precisely
future observations can constrain these parameters (\S
\ref{cosmology}).

In our calculations, we assume a cosmology with $\Omega_m=0.3$,
$\Omega_\Lambda=0.7$, $\Omega_b=0.046$, $H=100 \, h \, {\rm km \;
s^{-1} \; \Mpc^{-1}}$ (with $h=0.7$), $n=1$, and $\sigma_8=0.9$,
consistent with the most recent determinations \citep{spergel03}.  All
distances are measured in comoving coordinates.

\section{Velocity Field Effects}
\label{21cm}

The difference between the observed 21 cm brightness temperature at
the observed frequency $\nu$ and the CMB temperature today is
\citep{field59a}
\begin{equation}
T_b({\xvec}) = \frac{3 c^2 \, h_{\rm P} \, A_{1 0} \, n_H({\xvec}) \,
a^3\, \left[ T_s({\xvec}) - T_{\rm CMB}(z)\right]}{32 \pi \, k_B \,
T_s({\xvec}) \, \nu_0} \left| \frac{\partial r}{\partial \nu} \right|,
\label{eq:Tb21}
\end{equation}
where $c$ is the speed of light, $k_{\rm B}$ is the Boltzmann
constant, $h_{\rm P}$ is Planck's constant, $a = 1/(1 +z)$, $\bfx$ is
the spatial location, $T_{\rm CMB}$ is the the CMB temperature,
$A_{10} = 2.85 \times 10^{-15} ~s^{-1}$ is the spontaneous 21 cm
transition rate, $T_s$ is the spin temperature, $\nu_0 = 1420 \,
\MHz$, and $n_H$ is the number density of neutral hydrogen.  The
factor $|\partial r/\partial \nu|$ accounts for the Hubble flow as
well as peculiar velocities.  If we take the average of equation
(\ref{eq:Tb21}) we find
\begin{equation}
\bar{T}_b \approx 26 \, \bar{x}_H \left( \frac{\bar{T}_s - T_{\rm CMB}}{\bar{T}_s} \right)  \left(\frac{\Omega_b
h^2}{0.022} \right) \left( \frac{0.15}{\Omega_m h^2} \, \frac{1+z}{10}
\right) ^{1/2} ~{\rm mK}\, ,
\label{eq:Tavg}
\end{equation}
where $\bar{x}_H$ is the global neutral hydrogen fraction.
Fluctuations around $\bar{T}_b$ are at the tens of mK level on
megaparsec scales.

To calculate $\partial r/\partial \nu$, we relate comoving distance to
frequency \citep{bharadwaj04}
\begin{equation}
    r = \int_{\frac{\nu}{\nu_0 (1- v_r/c)}}^1 \frac{c \, da}{a^2 \, H(a)},
\end{equation}
where $v_r$ is the LOS peculiar velocity and $H(a)$ is the Hubble
constant.  Differentiating this expression, we find
\begin{equation}
 \frac{\partial r}{\partial \nu} = -\frac{c}{a^2 \nu_0 H}\,
 \left[1 - \frac{1}{H a}\frac{\partial v_r}{\partial r}
 \right],
\end{equation}
\noindent where we have dropped terms of order $[(Ha)^{-1} ~
\partial v_r /{\partial r}]^2$ and $v_r/c$.  In the limit $T_s \gg
T_{\rm CMB}$, fluctuations in the 21 cm brightness temperature at
${\xvec}$ can be expressed as
\begin{eqnarray}
\frac{\Delta \, T_b({\xvec})}{\tilde{T}_b}& = &\left(1 - \bar{x}_i
\,[1 + \delta_x({\xvec})]\right) \, \left(1 + \delta({\xvec})\right) \nonumber \\
&\times & \left(1
- \frac{1}{H a}\frac{\partial v_r({\xvec})}{\partial r}\right) -
\bar{x}_H, \label{dTeqn}
\end{eqnarray}
where $\bar{x}_i \equiv 1- \bar{x}_H$ is the global ionized fraction,
$\delta_x$ is the overdensity in the ionized fraction and $\delta$ is
the dark matter overdensity (at the scales and redshifts of interest,
the baryons trace the dark matter).  We define the normalized
temperature as $\tilde{T}_b \equiv \bar{T}_b/\bar{x}_H$.  In Fourier
space, since the linear theory velocity at redshifts where dark energy
is unimportant is $v(k, z) = -{\it i}\, H \, a \, {\kvec} \,
\delta_L/k^2$, the peculiar velocity term is $ \delta_{v} \equiv (H
a)^{-1} ~ \partial v_r/\partial r = - \mu^2 \delta_L$ where $\mu
\equiv \hat{{\kvec}}\cdot \hat{{\bf n}}$, the cosine of the angle
between the wavevector and the LOS, and $L$ denotes the linear
theory value.\footnote{The velocity field at $z \sim 10$ is in the
linear regime for $k \la 5 ~\Mpc^{-1}$.  See \citet{wang05b} for a
discussion of the effect of the non-linear velocity field on the 21 cm
signal.  Upcoming interferometers are most sensitive to scales at which
the velocity field is linear. }  If we keep terms to second order in
$\{\delta ,\delta^L\}$, the brightness temperature power spectrum is
\begin{eqnarray}
\tilde{T}_b^{-2}~P_{\Delta T}({\bfk}) & = & \left[\bar{x}_H^2 \,
P_{\delta \delta} + P_{x x} - 2 \bar{x}_H \, P_{x \delta} + P_{x
\delta x \delta} \right] \nonumber \\
& + & 2 \mu^2\left[ \bar{x}_H^2 \, P_{\delta_L
\delta}- \bar{x}_H \, P_{x \delta_L}\right] + \mu^4 \left[ \bar{x}_H^2
\, P_{\delta_L \delta_L} \right] \nonumber \\
& + & \left[2 P_{x \, \delta
\, \delta_v x} + P_{x \, \delta_v \, \delta_v \, x}\right],
\label{pseqn}
\end{eqnarray}
and we note that $P_{xx} = \bar{x}_i^2 \, P_{\delta_x \delta_x}$ and
$P_{x \delta} = \bar{x}_i \, P_{\delta_x \delta}$.  In our
calculations, we drop the connected part and set $P_{x \delta x
\delta} = P_{x \delta}^2 + P_{x x} \, P_{\delta \delta}$.  In
equation \ref{pseqn}, we have decomposed the power spectrum into
powers of $\mu$; the last bracket in this decomposition has a
non-trivial dependence on $\mu$. For notational convenience, we
refer to the k-dependent coefficients in equation (\ref{pseqn}) as
$P_{\mu^0}$, $P_{\mu^2}$ and $P_{\mu^4}$ and the terms in the last
bracket as $P_{f(\mu, k)}$. \citet{barkana04a} argue that the above
decomposition should allow one to extract the ``physics'' -- ($P_{\delta_L \delta_L}$)-- from the ``astrophysics''-- ($P_{xx}$ and
$P_{x \delta}$).  The terms in the last bracket in equation
(\ref{pseqn}) were omitted in their analysis, but must be included
if reionization is patchy because $\delta_x \sim 1$ on scales at or
below the bubble size.  If we drop the connected fourth moments, the
$P_{f(\mu, k)}$ terms are given by\footnote{These equations are
exact if fluctuations are Gaussian.}
\begin{eqnarray}
P_{x \, \delta \, \delta_v \, x}({\bfk}) & =& \int \frac{d^3{\bfk'}}{(2 \pi)^3} (\nhat \cdot {\hat{\bfk'}})^2 \, [P_{x
\delta_L}(k') ~P_{x \delta}(|{\bfk - \bfk'}|) \nonumber \\
& + & P_{\delta_L \delta}(k')~ P_{x x}(|{\bfk - \bfk'}|) ], \label{bad1eqn} \nonumber \\
P_{x \, \delta_v \, \delta_v \, x}({\bfk})& =& \int \frac{d^3{\bfk'}}{(2 \pi)^3} [(\nhat \cdot {\hat{\bfk'}})^4 \, P_{\delta_L \delta_L}(k') ~P_{x x}(|{\bfk - \bfk'}|) \nonumber \\
&+&(\nhat \cdot {\hat{\bfk'}})^2 ~ \left(\nhat\cdot \left[\frac{\bfk - \bfk'}{|{\bfk - \bfk'}|}\right] \right)^2 \nonumber \\
& \times & P_{x \delta_L}(k') ~P_{x \delta_L}(|{\bfk - \bfk'}|)],
\label{bad2eqn}
\end{eqnarray}
\noindent where $k' = ( \kvec' \cdot \kvec')^{1/2}$. These terms can
contaminate the $\mu$ decomposition.  On large scales, $P_{f(\mu,k)}$
does not depend on $\mu$ and therefore will contaminate only
measurements of $P_{\mu^0}$.  As we go to progressively smaller
scales, $P_{f(\mu,k)}$ contributes power to higher order terms in
$\mu$.

The evolution of the ionized fraction over a mode can also affect the
spherical symmetry of $P_{\Delta T}$, since time is changing in the
LOS direction but not in the angular directions. The magnitude of
this effect depends strongly on the morphology of reionization and is
discussed in Appendix \ref{evolution}.

\subsection{Model}

To model reionizaton by stellar sources, simulations must resolve
halos at least as small as the HI cooling mass ($\sim 10^8~
M_{\odot}$) as well as have boxes large enough to sample the size
distribution of HII regions, which can each reach larger scales
than $10 ~\Mpc$ \citep{furlanetto04a, barkana05}.  Recent simulations
have made significant strides towards attaining these goals
\citep{iliev05, kohler05}.  For the time being, semi-analytic models
of this epoch provide a convenient approach for modeling reionization
on the large scales that are relevant for upcoming observations.  In
this paper, we employ the physically motivated semi-analytic model
described in Furlanetto et al. (2004a; hereafter FZH04) to calculate
$P_{\Delta T}$.

Recent numerical simulations (e.g., \citealt{sokasian03, sokasian04,
ciardi03-sim}) show that reionization proceeds ``inside-out'' from
high density clusters of sources to voids, at least when the sources
resemble star-forming galaxies (e.g., Springel \& Hernquist 2003;
Hernquist \& Springel 2003).  We therefore associate HII regions with
large-scale overdensities.  We assume that a galaxy of mass $m_{\rm
gal}$ can ionize a mass $\zeta m_{\rm gal}$, where $\zeta$ is a
constant that depends on: the efficiency of ionizing photon
production, the escape fraction, the star formation efficiency, and
the number of recombinations.  Values of $\zeta \la 10-40$ are
reasonable for normal star formation, but very massive stars can
increase the efficiency by an order of magnitude \citep{bromm-vms}.

The criterion for a region to be ionized by the galaxies contained
inside it is then $f_{\rm coll} > \zeta^{-1}$, where $f_{\rm coll}$ is
the fraction of mass bound to halos above some minimum mass $m_{\rm
min}$.  We assume that this minimum mass corresponds to a virial
temperature of $10^4 {\rm K}$, at which point hydrogen line cooling
becomes efficient.  The function $f_{\rm coll}$ depends on the assumed
halo mass function.  \citet{furl-models} find that the choice of the mass
function has an insignificant effect on the FZH04 model. Here we use
the Press-Schechter mass function.  In the extended Press-Schechter
model \citep{bond91,lacey} the collapse fraction of halos above the
critical mass $m_{\rm min}$ in a region of mean overdensity $\delta_m$
is
\begin{equation}
f_{\rm coll} = {\rm erfc} \left( \frac{\delta_c - \delta_m}{\sqrt{2 \left[ \, \sigma^2_{\rm min} - \sigma^2(m, z) \right]}} \right),
\end{equation}
where $\sigma^2(m, z)$ is the variance of density fluctuations on the
scale $m$, $\sigma^2_{\rm min} \equiv \sigma^2(m_{\rm min}, z)$ and
$\delta_c \approx 1.686$, the critical density for collapse.  With this
equation for the collapse fraction, we can write a condition on the
mean overdensity within an ionized region of mass $m$,
\begin{equation}
 \delta_m \ge
\delta_B(m,z) \equiv \delta_c - \sqrt{2} K(\zeta) [\sigma^2_{\rm min}
- \sigma^2(m, z)]^{1/2},
\label{eq:deltax}
\end{equation}
where $K(\zeta) = {\rm erf}^{-1}\left(1 - \zeta^{-1} \right)$.

FZH04 showed how to construct the mass function of HII regions
from $\delta_B$ in a manner analogous to that of the halo mass function
\citep{press,bond91}.  The barrier in equation (\ref{eq:deltax}) is
well approximated by a linear function in $\sigma^2$, $\delta_B
\approx B(m) = B_0 + B_1 \sigma^2(m)$, where the redshift dependence
is implicit. In that case, the mass function has an analytic expression
\citep{sheth98}:
\begin{equation}
n(m) = \sqrt{\frac{2}{\pi}} \
\frac{\bar{\rho}}{m^2} \ \left| \frac{d \ln \sigma}{d \ln m}
\right| \ \frac{B_0}{\sigma(m)} \exp \left[ - \frac{B^2(m)}{2
\sigma^2(m)} \right],
\label{eq:dndm}
\end{equation}
 where $\bar\rho$ is the mean density of the Universe.  Equation
(\ref{eq:dndm}) gives the comoving number density of HII
regions with masses in the range $(m,m+dm)$.  The crucial difference
between this formula and the standard Press-Schechter mass function
occurs because $\delta_B$ is a decreasing function of $m$. The
barrier becomes more difficult to cross on smaller scales,
which gives the bubbles a characteristic size.

We can calculate $P_{xx}$ and $P_{x \delta}$ for the FZH04 model with
the semi-analytic method described in \citet{mcquinn05}. It is more
difficult to calculate $P_{x \delta_L}$.  \citet{mcquinn05} showed
that it is not necessary to consider bubble substructure in this
analytic model when constructing $P_{x \delta}$; -- only the size of a
bubble and the average density within a bubble are important.  Since we can
ignore bubble substructure, this implies that $P_{x \delta}
\rightarrow P_{x \delta_L}$ when the effective HII bubble radius
reaches linear scales.  This happens in the FZH04 model when
$\bar{x}_i \gtrsim 0.25$.  In the opposite limit, when the bubbles are
very small, this term is subdominant to density fluctuations and so
has a small effect on the power spectrum.  We set $P_{x \delta_L}=
P_{x \delta}$ for all times.  This will result in an overestimate of
this term when $\bar{x}_i$ is small and therefore an underestimate of
$P_{\mu^2}$.

For our calculations in this section, our objective is not to model
the 21 cm power spectrum for different reionization scenarios and
discuss morphological differences.  In the context of the FZH04 model,
this has been done in FZH04, \citet{furlanetto04d}, and
\citet{furl-models}.  Instead, we restrict ourselves to one
parameterization of this model, setting $\zeta(z) = 12$, in order to
illustrate the effect of redshift-space distortions on $P_{\Delta T}$.
For this parameterization, the EOR spans roughly the redshifts $8 -
15$.  In reality, $\zeta$ will have some time dependence, and it may
even have a very complicated evolution.  Fortunately, we find that the
parameterization $\zeta = 12$ is representative of the FZH04 model:
while varying the function $\zeta(z)$ will change $\bar{x}_i(z)$, if
we identify the same ionization fraction for different values of $\zeta$, the
values of $P_{xx}$ are quite similar (FZH04).

Of course, reionization might proceed differently than in this
analytic model. The parameter $\zeta$ may depend on the mass of the
dark matter halo. This can have a modest effect on the
characteristic size of the bubbles and the large-scale bubble bias
\citep{furl-models}. Also, recombinations might play a larger role
in shaping the morphology of reionization. Employing reasonable
assumptions for the clumpiness of the IGM, \citet{furlanetto05} show
that the effect of recombinations in the FZH04 model is only
important at $\bar{x}_i \ga 0.7$, increasing in importance as
$\bar{x}_i \rightarrow 1$. The presence of mini-halos may make
recombinations more important \citep{shapiro03}, but recent
work suggests that this may not be the case \citep{ciardi05}. So
long as recombinations are not important, \citet{furl-models} showed
that any model where the ionizing sources trace the collapsed
regions must have a qualitatively similar bubble size distribution
to the FZH04 distribution.

\begin{figure}
\epsfig{file=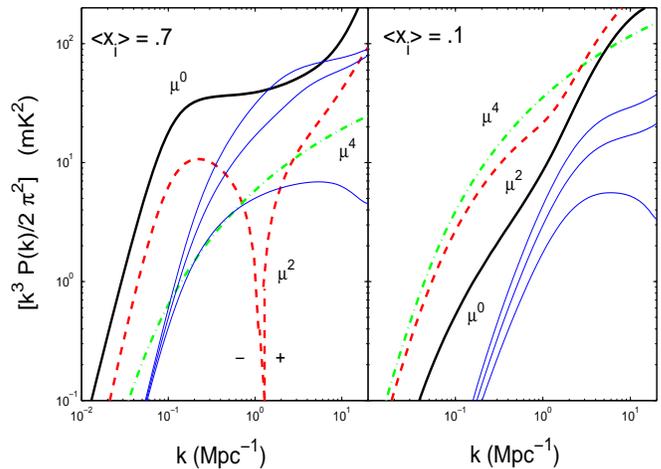, width=8.7cm, height = 6.2cm}
\caption{The $\mu$ decomposition of the signal (see equation
\ref{pseqn}) for $\bar{x}_i = 0.1$ and $0.7$, corresponding to $z =
13.5$ and $9$ in the $\zeta = 12$ model.  The thick solid, thick dashed and
thick dot-dashed curves show $P_{\mu^0}$, $P_{\mu^2}$, and $P_{\mu^4}$,
respectively.  The three thin solid curves show $P_{f(\mu, k)}$,
calculated using equation (\ref{bad2eqn}) with $\mu^2 = 0.0, 0.5$ and
$1.0$ (in order of increasing amplitude).}
\label{pscompfig}
\end{figure}

Figure \ref{pscompfig} plots for $\zeta = 12$ the components of the
dimensionless power spectrum, $k^3 \, P_{\Delta T}/2 \pi^2$, that have
different $\mu$ dependences. The thick solid, thick dashed and thick
dot-dashed curves indicate the $P_{\mu^0}$, $P_{\mu^2}$, and
$P_{\mu^4}$ terms, respectively.  The $\mu$ dependence of $P_{f(\mu,
k)}$ is nontrivial.  The three thin solid curves indicate the total
contribution from $P_{f(\mu, k)}$ for $\mu^2 = 0.0, 0.5$ and $1.0$ (in
order of increasing amplitude).

At $\bar{x}_i = 0.1$, $P_{\mu^4}$ is the largest of the $\mu$-terms
(see Fig. \ref{pscompfig}, {\it right}).  This is because the neutral
regions are underdense on average, and, at small $\bar{x}_i$, this
results in a suppression of $P_{\mu^0}$ and $P_{\mu^2}$.  The $\mu$
decomposition of the signal can be much different at larger ionization
fractions.  For $\bar{x}_i = 0.7$, $P_{\mu^0}$ is much larger than the
other terms at most scales (see Fig.\ref{pscompfig}, {\it left}), and
$P_{\mu^2}$ becomes negative at small $k$ because the neutral gas is
underdense on average.  Finally, at smaller scales than the bubbles
size, $P_{f(\mu, k)}$ is larger than $P_{\mu^2}$ and $P_{\mu^4}$, and
is even larger than $P_{\mu^0}$ at $k \approx 3 \, \Mpc^{-1}$.

\begin{figure}
\epsfig{file=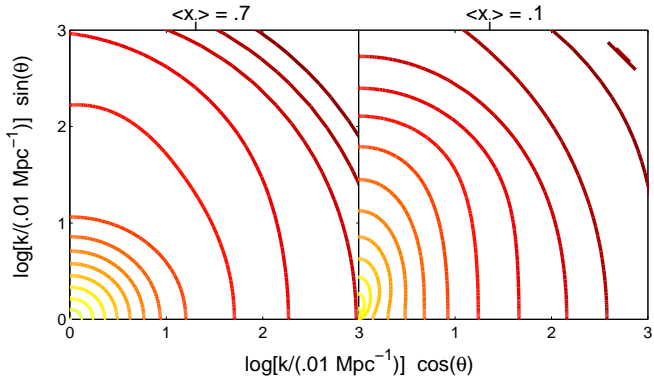, width=8.7cm}
\caption{Contours of constant $k^3 P_{\Delta T}(k)$ for the same
signal as in Figure \ref{pscompfig}. The horizontal axis is the LOS
direction, and the vertical axis is the transverse direction.  The
coordinate transformation $(k_{\perp}, k_{||}) \rightarrow \log(k/(0.01
~\Mpc^{-1})) (\sin(\theta), \cos(\theta))$ preserves circles of
constant power.}
\label{fig:contours}
\end{figure}

The evolution of the spherical symmetry as a function of $\bar{x}_i$
is non-trivial.  When the ionization fraction is small, the
redshift space distortions are important on all scales.  In the
opposite limit, when the ionization fraction is large, the
redshift-space effects are less important since the bubble-bubble term
$P_{xx}$, which enters through $P_{\mu^0}$, dominates the signal
(Fig. \ref{pscompfig}, {\it left}).  The evolution of the angular
symmetry of the signal is illustrated in Figure \ref{fig:contours},
which plots the contours of constant $k^3 \, P_{\Delta T}(k)$ for
$\bar{x}_i = 0.1$ and $\bar{x}_i = 0.7$ ({\it right and left panels,
respectively}).  When $\bar{x}_i = 0.7$, the signal is fairly symmetric
at smaller $k$-modes than the bubble scale.  At larger values of $k$ or
at small ionization fractions, density fluctuations dominate the
signal, and the power spectrum can be very asymmetric.  Because of
this, it may be possible to determine
the characteristic bubble size by observing the angular dependence of $P_{\Delta T}$.

\begin{figure}
\epsfig{file=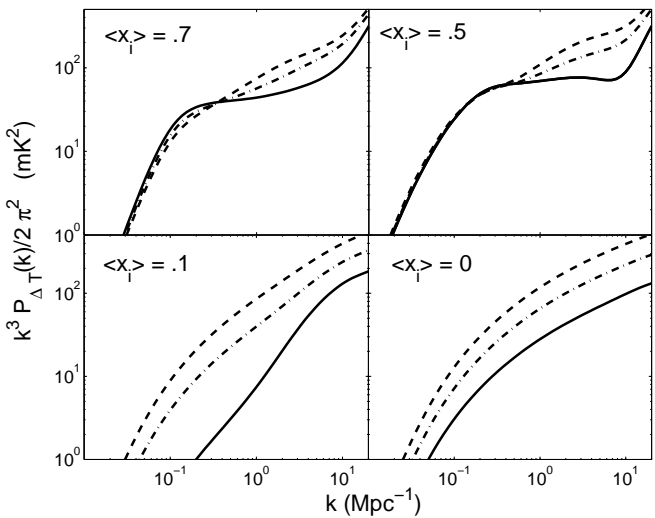, width=8.7cm}
\caption{Dimensionless power spectrum $k^3
\,P_{\Delta T}/2 \pi^2$ for $\mu^2$ equal to $0.0, \, 0.5,$ and $1.0$
({\it solid, dash-dot, and dashed lines, respectively}) for four times during
reionization in the $\zeta= 12$ model, corresponding to $z =$ 9, 9.8,
13.5 and 20 (in order of decreasing $\bar{x}_i$).  The signal is most asymmetric at scales where density fluctuations dominate.} \label{psfig}
\end{figure}

In Figure \ref{psfig}, we plot the signal $[k^3 \, P_{\Delta
T}(k,  \mu)/2 \pi^2]$ for modes with $\mu^2 = 0.0, 0.5, ~{\rm and} ~
1.0$ ({\it the solid, dot-dashed, and dashed curves, respectively}) at four times during reionization.  Between $\bar{x}_i \sim 0.0 - 0.1$,
density fluctuations dominate the signal and the signal is very
asymmetric.  By $\bar{x}_i = 0.5$, the neutral fraction
fluctuations contribute most of the power on large scales.  When these
fluctuation dominate, the 21 cm power spectrum develops a ``shoulder''
on scales near the characteristic bubble size.  This feature moves to
larger scales as the bubbles grow.

In \S \ref{detectors}, we show that upcoming interferometers will be
more sensitive to modes with certain orientations relative to the
LOS.  It turns out that arrays that are very compact (i.e. have most
of their antennae within a 1 km region), such as MWA, are most
sensitive to $k$-modes that are oriented along the LOS.  The fact
that these modes have more power enhances MWA's sensitivity.
Conversely, it is difficult to separate the terms $P_{\mu^0}$,
$P_{\mu^2}$ and $P_{\mu^4}$ with observations that are most sensitive
to the modes along the LOS.

\section{Sensitivity to the 21 cm Power Spectrum}
\label{noise}

In this section, we summarize how to calculate the variance on
interferometric observations of the 3-D 21 cm power spectrum.  Our
calculation follows that of \citet{morales05}, and extends their
calculation to capture the angular dependence of the 3-D signal.
\citet{white99} and \citet{zald04} do a similar interferometric
detector noise calculation, but for the angular power spectrum.  For
21 cm observations, the 3-D power spectrum is more interesting than
the angular power spectrum owing in part to the the $\mu$ dependence
of the signal.  In addition, the 3-D power spectrum will allow us to
measure many more independent modes.

The 21 cm signal will be observed with radio interferometers, which
measure visibilities.  The visibility for a pair of
antennae, quantified as a temperature, is given by
\begin{equation}
V(u, v, \nu) =  \int{d{\nhat}
\, \, \Delta T_b(\nhat, \nu) \, A_\nu(\nhat) e^{2 \pi {\it
i}{u \choose v} \cdot \nhat}},
\label{viseqn}
\end{equation}
where $(u, v)$ is a vector that gives the number of wavelengths at
frequency $\nu$ between the antennae and $A_\nu(\nhat)$ is the
contribution to the primary beam in the direction $\nhat$.  Here, we
are working in the flat sky approximation; this is adequate since
upcoming experiments are most sensitive to angular modes with
wavelength $\theta < 0.1$ radians.

We assume that the visibilities are complex Gaussian random variables,
such that the likelihood function of the covariance matrix $C_{i\,j} =
\langle {V}^*_i \, {V}_j \rangle$ for $n$ visibilities, where the
asterisk indicates a complex conjugate, is
\begin{equation}
{\cal L}({\boldsymbol{C}}) = \frac{1}{\pi^n \, {\rm
det}{\boldsymbol{C}}} ~ \exp \left[- \mathop{\sum}_{i,j}^n{\,{
V}_i^* \, C_{ij}^{-1} \, {V}_j} \right].
\label{eq:likelyhood}
\end{equation}
Because the visibilities are complex and ${V}(u, v, \nu) = {V}(-u, -v,
\nu)^*$, when counting the number of independent $u$-$v$ pixels we
restrict ourselves to the half space.  In this section, it will be
more convenient to work with the Fourier transform of $\boldsymbol{V}$
in the frequency direction.  This is just a change of basis of
$\boldsymbol{V}$ and $\boldsymbol{C}$ in equation
(\ref{eq:likelyhood}).

For upcoming arrays, $\boldsymbol{C}$ will be dominated by the detector
noise on most scales.  The RMS detector noise fluctuation per
visibility of an antennae pair after observing for a time $t_0$ in
one frequency channel is \citep{rohlfs}
\begin{equation}
\Delta V^N = \frac{\lambda^2 T_{\rm sys}}{ A_e \, \sqrt{\Delta
\nu \,t_0}}, \label{vrmseqn}
\end{equation}
where $T_{\rm sys}(\nu)$ is the total system temperature, $A_e$ is the
effective area of an antenna, and $\Delta \nu$ is the width of the
frequency channel.

For an observation with bandwidth $B$, where $B \gg \Delta \nu$, if we
Fourier transform the observed visibilities in the frequency
direction, we then have a 3-D map of $\tilde{I}(\uvec) \equiv \int
d\nu {V}(u, v, \nu) \exp[2 \pi i \nu \eta]$, in which $\uvec = u \,
\hat{\boldsymbol{i}} +v \, \hat{\boldsymbol{j}} + \eta \,
\hat{\boldsymbol{k}}$ and $\eta$ has dimensions of time.  If we
perform this transform on just the detector noise component
$\boldsymbol{V}^N$ of the visibility map $\boldsymbol{V}$, we have
\begin{eqnarray}
\tilde{I}^N(\bfu) &=& \mathop{\sum}_{i = 1}^{B/\Delta \nu} \,
{V}^N(u, v, \nu_i) \, \exp[2 \pi i \nu_i \eta]\, \Delta \nu \\
& =&\mathop{\sum}_{i=1}^{B/\Delta \nu} \, {V'}^N(u, v, \nu_i) \, \Delta \nu,
\end{eqnarray}
where we have absorbed $\exp[2 \pi i \nu \eta]$ into a new variable
$V'^N$ that has the same RMS as ${V}^N$ and the frequency channels
$\nu_1,..., \nu_{B/\Delta \nu}$ are spaced $\Delta \nu$ apart.  It
follows that the detector noise covariance matrix for a single
baseline is \citep{morales05}
\begin{eqnarray}
C^N_{1 b}(\uvec_i, \uvec_j) & = & \langle \tilde{I}^N(\uvec_i)^* \,  \tilde{I}^N( \uvec_j) \rangle \nonumber \\
& = & \langle \left(\mathop{\sum}_{m = 1}^{B/\Delta \nu} \, {V'}^N(\uvec_i, \nu_m) \, \Delta \nu \right)^* \nonumber \\
& & \times \left(\mathop{\sum}_{l = 1}^{B/\Delta \nu} \, {V'}^N(\uvec_j, \nu_l) \, \Delta \nu\right) \rangle \nonumber \\
& = & B \, \Delta \nu \, \left(\Delta V^N\right)^2 \, \delta_{ij} \label{eq:Icorr2}\\
& = & \left(\frac{\lambda^2 \, B \, T_{\rm sys}}{
A_e}\right)^2 \, \frac{\delta_{ij}}{B \, t_0}.
\label{eq:Icorr}
\end{eqnarray}
To reach equation (\ref{eq:Icorr2}), we note that different
 ${V}'^N(\uvec_i, \nu_m)$ are uncorrelated, and equation
 (\ref{eq:Icorr}) follows from equations (\ref{vrmseqn}) and
 (\ref{eq:Icorr2}).  Note that equation (\ref{eq:Icorr}) only depends
 on $B$ and not on $\Delta \nu$: finer frequency resolution comes at
 no added cost.

We now estimate the average observing time $t_\kvec$ for an array of
 antennae to observe a mode $\kvec$ as a function of the total
 observing time $t_0$ (note that there is an isomorphism $\uvec
 \leftrightarrow \kvec$ and that $2 \pi \uvec_{\perp}/x =
 \kvec_{\perp}$, where x is the conformal distance to the emission).
 At any time, the number density of baselines that can observe the
 mode $\uvec$ is $n(\uvec_{\perp})$.  We assume this density is
 rotationally invariant and define $\theta$ to be the angle between
 $\kvec$ and the LOS. Integrating $n(\uvec_{\perp}) \, du \,dv$ over
 the half plane yields $N_{\rm base} \approx N_{\rm ant}^2/2$,
 where $N_{\rm ant}$ is the number of antennae.  Since the telescope
 observes a region in the $u$-$v$ plane equal to $\delta u \delta v
 \approx A_e/\lambda^2$, each visibility is observed for a time
\begin{equation}
t_\kvec \approx \frac{A_e \, t_0}{\lambda^2}\, n(x \, |\kvec| \,{\rm
sin}(\theta)/ 2 \pi), \label{time}
\end{equation}
where $t_0$ is the total observing time for the interferometer.  It
follows that the detector noise covariance matrix for an
interferometer is
\begin{equation}
C^N(\bfk_i, \bfk_j) = \left(\frac{\lambda^2 \, B \, T_{\rm sys}}{
A_e}\right)^2 \, \frac{\delta_{i j}}{B \, t_{\bfk_i}}.
\label{eq:CN}
\end{equation}
(From now on we will use $\bfk$ rather than $\bfu$ to index
elements in $\boldsymbol{C}$).

We also want an expression for the contribution to $\boldsymbol{C}$ owing
to sample variance.  For a 3-D window function ${W}(\nhat,\nu) =
A_\nu(\nhat)~ F_\nhat(\nu)$, if we assume that different pixels indexed
by $\uvec$ are uncorrelated, the covariance matrix of the 21 cm
signal $\tilde{I}^{21}$ is
\begin{eqnarray}
C^{SV}(\kvec_i, \kvec_j) & = & \langle \tilde{I}^{21}(\uvec_i)^* \tilde{I}^{21}(\uvec_j) \rangle \nonumber \\
& \approx & ~\delta_{i j} ~ \int d^3\uvec' \, |\tilde{W}(\uvec_i - \uvec')|^2
    \, P^{21}_{\Delta T}(\uvec'),
\end{eqnarray}
where we have used that $\langle \Delta T^{21}(\bfu') \Delta
\,T^{21}(\bfu)\rangle = P_{\Delta T}(\bfu) ~ \delta^3(\bfu' -\bfu)$
and the definition of visibility (eqn. \ref{viseqn}).  We can simplify
$\boldsymbol{C}^{SV}$ further:
\begin{eqnarray}
C^{SV}(\kvec_i, \kvec_j)& \approx& ~ P^{\rm 21}_{\Delta
    T}(\uvec_i) \, \frac{\lambda^2 \, B }{A_e} ~ \delta_{i j} \label{correqn2}\\
& \approx& ~ P^{21}_{\Delta
    T}(\kvec_i) \, \frac{\lambda^2 \, B^2 }{A_e \, x^2 \, y} ~ \delta_{i j},
\label{correqn}
\end{eqnarray}
where to get to equation (\ref{correqn2}) we pull $P(\bfu_i)$ out of
the integral and use the fact that $\tilde{W}(\uvec)$ is different
from zero in an area $\delta u \delta v \delta \eta \approx
A_e/(\lambda^2 \, B)$ and must integrate to unity within the beam,
such that $\int d^3\uvec' \, |\tilde{W}(\uvec -\uvec')|^2 \approx
(\delta u \delta v \delta \eta)^{-1}$.  Equation (\ref{correqn2}) is
accurate for values of $|\uvec_i|$ much greater than the FWHM of
$\tilde{W}$.  The additional factor of $x^2 y/B$ that arises in
equation (\ref{correqn}) is because with our Fourier conventions
$P_{\Delta T}(\kvec) = x^2 y/B \,P_{\Delta T}(\uvec)$, where $y$ is the
conformal width of the observation.

Over the course of an observation, a large number of independent
Fourier cells will be observed in a region of real space volume ${\rm
Vol} = x^2 \, y \, \lambda^2/A_e$.  We have seen that the 21 cm power
spectrum is not spherically symmetric, but is symmetric around the
polar angle $\phi$.  Because of this symmetry, we want to sum all the
Fourier cells in an annulus of constant $(k, \theta)$ with radial
width $\Delta k$ and angular width $\Delta \theta$ for a statistical
detection.  The number of independent cells in such an annulus is
\begin{equation}
N_c( k, \theta) = 2 \pi {k}^2 \, \sin(\theta) \, \Delta k
\, \Delta \theta \times \frac{\rm Vol}{(2 \pi)^3}.
\label{cells}\end{equation}
Here, $(2 \pi)^3/{\rm Vol}$ is the resolution in Fourier space.  For
our calculations, we use equation (\ref{cells}) for $N_c$ when the
wavelength corresponding to $\bfk$ fits within the survey volume
(i.e. when $2 \pi/k \, \cos(\theta) < y$) and otherwise we set $N_c
=0$.\footnote{To more accurately capture these modes, we could
discretize $k$ and physically count the number of modes within the
volume.  However, our approximation is only inaccurate for small $k$.
Foregrounds will eliminate our ability to measure these
long-wavelength modes such that a more precise treatment is
unnecessary (\S \ref{sens_results}).}

When we sum equations (\ref{eq:CN}) and (\ref{correqn}) to get
 $\boldsymbol{C}$, the error in $P^{21}_{\Delta T} (\bfk)$ from a
 measurement in an annulus with $N_c(k, \theta)$ pixels
 is\footnote{The reader may be familiar with expressions for the error
 that contain a factor of $\sqrt{{2 }/{N_c}}$.  We do not have a factor
 of $\sqrt{2}$ in equation (\ref{eq:dP}) because each pixel has both a
 real an imaginary component.  Since we only count pixels in the half
 space, this formulation is equivalent.}
\begin{equation}
\delta P^{21}_{\Delta T} (k, \theta) \approx \sqrt{\frac{1}{N_c}}\,
\frac{ A_e \, x^2 \, y }{ \lambda^2 \, B^2} \,
\left[{\rm C}^{SV}(k, \theta) + {\rm C}^N(k, \theta) \right],
\label{eq:dP}
\end{equation}
where we have defined ${\rm C}(\bfk) \equiv C(\bfk, \bfk)$.  One can
trivially derive equation (\ref{eq:dP}) by calculating the Fisher
matrix $\boldsymbol{F}$ of the $P(\bfk_i)$ from $\boldsymbol{C}$
(e.g. eqn. \ref{eq:fishf2}, with $\tilde{\bfC} \rightarrow \boldsymbol{C}$),
noting that there are $N_c$ measurements of $P(k, \theta)$ and that
$\delta P^{21}_{\Delta T} (\bfk) = \sqrt{{F^{-1}}_{\bfk \bfk}}$.

Given a model for the data, the 1-$\sigma$ errors in the model parameters
$\lambda_i$ are ($\sqrt{{F'^{-1}}_{ii}}$), where
\begin{equation}
F'_{i j} = \sum_{\rm pixels} \frac{1}{(\delta P_{\Delta T})^2} \, \frac{\partial P_{\Delta T}}{\partial \lambda_i} \, \frac{\partial P_{\Delta T}}{\partial \lambda_j}.
\end{equation}
[See Appendix \ref{Ap:spav} for a useful formula for $\delta P_{\Delta
T}(k)$, the error in the angular-averaged power spectrum.]

In the next section, we extend the above analysis to include
foregrounds.  The calculation in this section assumes Gaussian
statistics, but the ionization fraction fluctuations on the
scale of the HII bubbles will not be Gaussian.  Numerical simulations
are necessary to quantify the degree of non-Gaussianity introduced by
patchy reionization.

\section{Foregrounds}
\label{foregrounds} The foregrounds at $\lambda = 21\, (1 +z)$ cm
will be at temperatures of hundreds to thousands of kelvin,
approximately $4$ orders of magnitude greater than the 21 cm
signal.  All the significant foreground contaminants should have
smooth power-law spectra.  Known sources of radio
recombination lines are estimated to contribute to the fluctuations
at an insignificant level \citep{oh03}.

Before fitting a model to the cosmological signal, it is necessary to clean
the foregrounds from the data.  The idea is to
subtract out a smooth function from the total signal prior to the
parameter fitting stage \citep{tegmark00}. Such
pre-processing is common with CMB data sets, and \citet{wang05} showed
that this procedure can also be used in handling 21 cm observations.

At the frequency $\nu_i$ in a pixel
with angular index $\bfk_\perp$, an interferometer measures
\begin{equation}
x_i = s(\bfk_\perp, \nu_i) + n(\bfk_\perp, \nu_i) + f(\bfk_\perp, \nu_i),
\label{eq:comp}
\end{equation}
where $s$ is the 21 cm signal in visibilities, $n$ is the detector
noise fluctuation, and $f$ is the foreground amplitude (all of which
are complex).  We will subsequently write the quantities $x, s, n$ and
$f$ measured at the N frequencies ${\bfnu} = (\nu_1, ..., \nu_N)$ with
resolution $\Delta \nu$ as the vectors ${\bfx}, {\bfs}, {\bfn}$, and
${\bff}$.  There is one key difference between our calculation and
that of \citet{wang05}.  Rather than subtracting from $\log({x})$ a
polynomial in $\log({\nu})$, which is functionally very similar to the
known foregrounds, we instead subtract a polynomial in ${\nu}$ from
$x$.  While this difference may require a higher order function to
adequately fit the data, it also permits an analytic treatment.

Fitting an order-${\rm n}$ polynomial to the vector $\bfx$ is
equivalent to projecting out the Legendre polynomials ${\bfP}_{0},{\bfP}_{1}, ..., {\bfP}_{\rm n}$, normalized such that
$\int_{0}^{B} P_{l_1} (2\nu/B - 1) P_{l_2}(2 \nu/B - 1) d\nu =
\delta_{l_1, l_2}$ and ${P_{l,\, i}} = P_{l} (2\nu_i/B - 1)\,
\sqrt{\Delta \nu}$.\footnote{The formalism discussed in this section
should apply to any complete set of orthogonal functions and not just
Legendre polynomials.}  Projecting out to order {\rm n}, our
cleaned signal is
\begin{equation}
\tilde{\bfx} = \left(1 - \sum_{l = 0}^{\rm n} \, {\bfP}_{l}\,{\bfP}_{l}^T
\right) \, \bfx ~\equiv~ \sum_{l = {\rm n} + 1}^N \,{\bfP}_{l}\,{\bfP}_{l}^T
\, \bfx.
\label{eq:tildex}
\end{equation}
and $\tilde{\bfs}$, $\tilde{\bfn}$ and $\tilde{\bff}$ are defined in
analogy to $\tilde{\bfx}$.  The covariance matrix for the cleaned
signal is $\tilde{\bfC} \equiv \langle \tilde{\bfx} \, \tilde{\bfx}^t
\rangle$.  Let us write ${\bf \Pi} = \sum_{l = {\rm n} + 1}^N
\,{\bfP}_{l}\,{\bfP}_{l}^T$.  We need to invert $\tilde{\bfC}$ to
calculate $\delta P^{21}_{\Delta T}$.  Because $\tilde{\bfC}$ is
singular, to invert $\tilde{\bfC}$ we use the trick $\tilde{\bfC}
\rightarrow \tilde{\bfC} + \eta \, \left(\sum_{l = 1}^{\rm n} \,
{\bfP}_{l}\,{\bfP}_{l}^T \right) \equiv \tilde{\bfC}_* $, where
$\eta$ is a large number.  This method for inverting $\tilde{\bfC}$
does not lose information \citep{tegmark97}.  In the basis of the
${\bfP}_{l}$,
\begin{eqnarray}
\tilde{\bfC}_* & = & \langle \tilde{\bfx} \,\tilde{\bfx}^\dag \rangle +
    \eta \, \sum_{l = 1}^{\rm n} \, {\bfP}_{l}\,{\bfP}_{l}^T \nonumber \\
&= & {T_N}^2 \, {\bf \Pi} +
    \tilde{\bff} \, \tilde{\bff}^\dag + \tilde{\bfs} \, \tilde{\bfs}^\dag + \eta \,
    \sum_{l = 1}^{\rm n} \, {\bfP}_{l}\,{\bfP}_{l}^T,
\end{eqnarray}
where $T_N$ is the $\Delta T_N$ defined in equation (\ref{vrmseqn}) except
with the replacement $t_0 \rightarrow t_{\bfk_{\perp}}$ ($\langle \bfn
\, \bfn^\dag\rangle$ is diagonal in the chosen basis).  When the
detector noise dominates over the signal, the inverse of $\tilde{\bfC}_*$ is
\begin{equation}
\tilde{\bfC}_*^{-1} \approx \frac{\bfI}{{T_N}^2} - \frac{\tilde{\bff} \, \tilde{\bff}^\dag}{{T_N}^2 \, ({T_N}^2 + \tilde{\bff}^2)},
\end{equation}
where $\bfI$ is the identity matrix.
Here, we have dropped terms proportional to $1/\eta$.  If the foregrounds
can be cleaned well below the signal,
the Fisher matrix for the 21 cm power spectrum is
\begin{eqnarray}
F_{k, k'} &= & \frac{\partial^2 ~ \langle \log{\cal L} \rangle}{\partial P^{\rm 21}_{\Delta T}(k)\, \partial P^{\rm 21}_{\Delta T}(k') }\label{eq:fishf} \\
&=& {\rm tr}\left[\tilde{\bfC}_*^{-1} \, \frac{ \partial \tilde{\bfC}}{\partial P^{\rm 21}_{\Delta T}(k)} \, \tilde{\bfC}_*^{-1}
\frac{ \partial \tilde{\bfC}}{\partial P^{\rm 21}_{\Delta T}(k')}
\right], \label{eq:fishf2}
\end{eqnarray}
where ${\cal L}$ is defined in equation (\ref{eq:likelyhood}) and only
visibilities with the same $\bfk_{\perp}$ are used.

We want to constrain the parameters $P_{\Delta T}(\bfk_\perp,
k_{||})$.  We can write $P_{\Delta T}(\bfk_\perp, k_{||})$ in terms of
the signal ${\bfs}$ via a Fourier transform:
\begin{equation}
\langle \bfs \bfs^\dag \rangle_{\bfk_\perp} = \sum_k \, w~ P^{\rm
21}_{\Delta T}(\bfk_\perp, k) \bfmu_k \bfmu_k^\dag.
\label{eq:ftPk}
\end{equation}
Note that $k$ in ${\bf \mu}_k$ and in $P_{\Delta T}( \bfk_{\perp}, k)$
denotes the LOS component of ${\bf \it k}$ rather than the norm of
${\bfk}$.  Here the Fourier vector $\bfmu_k \propto \exp[i y/B \, k \,
\bfnu]$, where $y$ is the length of the box, and $w \approx \lambda^2
B^2/(A_e \, x^2 \, y)$ (see equation \ref{correqn}).  It follows from
equation (\ref{eq:ftPk}) that
\begin{equation}
\frac{\partial \tilde{\bfC}}{\partial P^{\rm 21}_{\Delta T}(k)} = w \, {\bf \Pi} \, \bfmu_k \, \bfmu_k^\dag \, {\bf \Pi}^T = w \, \tilde{\bfmu}_k \, \tilde{\bfmu}_k^\dag.
\end{equation}
For the $k$ and $k'$ at which the foregrounds can be cleaned well below the
signal, the Fisher matrix is
\begin{equation}
F^{\bfk_{\perp}}_{k, k'} \approx w^2 \, \left[
\left(\frac{\tilde{\bfmu}_k^\dag \,
\tilde{\bfmu}_{k'}}{T_N^2}\right) - \frac{\left(\tilde{\bfmu}_k^\dag
\bff \right) \, \left(\bff^\dag \tilde{\bfmu}_{k'} \right)}{T_N^2 \,
\left(T_N^2 + \tilde{\bff}^2 \right)} \right]^2. \label{forfish}
\end{equation}
Since pixels with different $\bfk_\perp$ are
independent, we can combine the error in $P^{21}_{\Delta T}(k)$ from
all pixels with the same $(k, \theta)$ as in \S \ref{noise}.
Therefore, if cleaning is successful, the combined error from the
pixels in an annulus indexed by $(k, \theta)$ is
\begin{equation}
 \delta P^{21}_{\Delta T}(k_i, \theta) \approx \frac{1}{\sqrt{N_{\rm c}(k_i, \theta)}} ~\sqrt{\left({\bfF^{\bfk_{\perp}}}^{-1} \right)_{i, i}}.
\label{eq:forstat}
\end{equation}
This equation is a good approximation for $k$ at which $ \bfmu_k^\dag
\, \langle \tilde{\bfs} \, \tilde{\bfs}^\dag \rangle \, \bfmu_k \gg
\bfmu_k^\dag \, \tilde{\bff} \, \tilde{\bff}^\dag \, \bfmu_k $.  If we
use the approximate orthogonality of the $\tilde{\bfmu}_k$ (note that
the LOS $k$ are sampled $2\pi/y$ apart), this condition reduces
to
\begin{equation}
 P^{21}_{\Delta T}(\bfk_{\perp}, k)  \gg  \frac{\bfmu_k^\dag \, \tilde{\bff} \, \tilde{\bff}^\dag \, \bfmu_k}{ w \,\left(\tilde{\bfmu}_k^\dag \, \tilde{\bfmu}_k \right)^2} \equiv Q_{\bfk_{\perp}}(k, {\rm n}).
\label{eq:cond}
\end{equation}
The larger the bandwidth, the higher order polynomial it should take
to fit the data.  To optimize the foreground removal procedure, the
minimum ${\rm n}$ should be chosen such that the condition given in
(\ref{eq:cond}) is satisfied.  The larger the value of ${\rm n}$, the
more power will be removed from the 21 cm signal.

The formalism in this section can be easily generalized to include the
situation in which the foregrounds are removed over a larger bandwidth
than the bandwidth over which the 21 cm signal is extracted.
Increasing the bandwidth over which the foreground removal is performed
will improve an interferometer's sensitivity to the cosmological
signal (\S \ref{sens_results}).

\subsection{Foreground Model}
\label{fmodels}
The three major foreground contaminants are extragalactic point
sources, Galactic bremsstrahlung and Galactic synchrotron.  The
Galactic synchrotron comprises about 70\% of the foreground
\citep{shaver99}, but the extragalactic point sources may be the
hardest to remove \citep{dimatteo02}.  Here we are not concerned
with the overall amplitude of these foregrounds, since an
interferometer cannot measure the $\bfk_{\perp} = 0$ mode.

To model the angular power spectrum of the Galactic synchrotron, we employ
the function
\begin{equation}
\frac{l^2 \, C_l(\nu_1, \nu_2)}{2 \pi} =
\left(\frac{l}{l_0} \right)^{2-\beta} \, T^{\rm syn}_{l_0}(\nu_1) \, T^{\rm syn}_{l_0}(\nu_2),
\end{equation}
\begin{equation}
T^{\rm syn}_{l_0}(\nu) = A^{\rm syn}_{l_0} \, \left(\frac{\nu}{150
~\MHz}\right)^{-\alpha_{\rm syn} - \Delta \alpha_{\rm syn}\,
\log(\frac{\nu}{150~ \MHz})},
\end{equation}
$\alpha_{\rm syn} = 2.55$, $\Delta \alpha_{\rm syn} = 0.1$, $\beta =
2.5$ and $A^{\rm syn}_{l_0 = 5} =25~ K$ \citep{shaver99, tegmark00,
wang05}.  The latter two values are extrapolated from $30~ {\rm GHz}$
CMB observations.

  For the extragalactic point sources we employ the \citet{dimatteo02}
model.  \citet{gnedin04} points out that \citet{dimatteo02} make
very pessimistic parameter choices for this model.  As a result, this
model probably overestimates the contribution from extragalactic point
sources.  The extragalactic point source contribution has two
components, a Poisson component and a clustering component.  Bright
sources can be removed from the map prior to the foreground fitting
stage.  Once bright sources are cleaned, the Poisson component is
\begin{equation}
C_{l}^{\rm pois} = \int_0^{T_{\rm cut}} dT \int d\zeta  ~ T^2
\frac{\partial^2 N}{
\partial T \partial \zeta},
\end{equation}
where $T_{\rm cut}$ is the minimum brightness temperature of the
sources that can be cleaned and $\partial^2 N/ \partial T
\partial \zeta$ is the number of sources per unit brightness
temperature at $150~ \MHz$ per $\zeta$ -- the spectral index of a
source -- per steradian.

To model the clustering term, we assume that the spectral indexes of
sources are spatially uncorrelated and set the correlation function of
the extragalactic sources to be $w(\theta) =
[{\theta}/({\theta_*})]^{-\beta}$, such that
\begin{equation}
C_{l}^{\rm clust}(\nu_1, \nu_2) \propto  \left(l \right)^{\beta -2} ~
T_{eg}(\nu_1) \, T_{eg}(\nu_2),
\end{equation}
where $\beta = 0.85$, $\theta_* = 4'$ \citep{dimatteo02} and
\begin{equation}
T_{eg}(\nu) = \int_0^{T_{\rm cut}} dT_{\nu_0} \int d\zeta
\frac{\partial^2 N}{\partial T_{\nu_0} \partial \zeta} \,
\left[T_{\nu_0} (\frac{\nu}{\nu_0})^{-\zeta}\right].
\end{equation}

We model the probability distribution of the spectral index $\zeta$ as
a spatially constant Gaussian with standard deviation $\delta \zeta =
0.3$ and mean $\bar{\zeta} = 2.8$ \citep{tegmark00}.  We assume 4
sources sr$^{-1}$ mJy$^{-1}$ at $880$ mJy and a power-law scaling in
flux with exponent $-1.75$ \citep{dimatteo02}.  Furthermore, we take
$T_{\rm cut} = 7 \times T_{\rm inst}$ where the instrumental
sensitivity limit is
\begin{equation}
T_{\rm inst} \approx \frac{\lambda^2 \, T_{\rm sys}}{N_{\rm ant} \, A_e \, \sqrt{2 \, t_0 \, B}}.
\end{equation}
The values of $S_{\rm cut} = 6.9 \times 10^5 \, [T_{\rm cut}/({\rm 1
\,K})] \,[\nu/(150 \, \MHz)]^{2}$ mJy for MWA, LOFAR, and the Square
Kilometer Array (SKA) after $1000$ hr of observations at $150~
\MHz$ with $B = 6 ~\MHz$ are listed in Table \ref{table1}.

The foreground power is dominated by the Galactic synchrotron at most
scales.  Because of this, this foreground is the most difficult to
remove from the 21 cm map.  At $l \ga 5000$, the extragalactic point
sources fluctuations start to become important.  In this analysis, we
ignore the contribution owing to Galactic and extragalactic
bremsstrahlung emission.  The Galactic emission is expected to account
for roughly $1\%$ of the contamination at the relevant frequencies
\citep{shaver99} and contributes a negligible amount of power at all
scales.  While there is large uncertainty in the extragalactic
bremsstrahlung, its contribution will also be minor at the relevant scales
\citep{santos05}.

With this model for foregrounds, we can calculate the experimental
sensitivities using the formulas in the first part of this section if
we note that $\langle f(\nu_1)^* \, f(\nu_2) \rangle_{k_{\perp} = l/x}
= \lambda^2/A_e \, C_{l}(\nu_1, \nu_2)$, where the prefactor of
$\lambda^2/A_e$ comes from $\int dudv \, |\tilde{A}_\nu(u,v)|^2$ (see
eqn. \ref{correqn2}).

We have made several simplifying assumptions for the form of the
foregrounds.  For example, extragalactic point sources will not exactly
have a Gaussian distribution of spectral indexes and the
frequency dependence of the foregrounds may be a function of ${\it l}$.
While we anticipate that our simplifications will have a negligible
effect on the overall foreground cleaning, this is a question that is
beyond the scope of our analysis.  (See, e.g. \citet{santos05} for a
treatment of more complicated foreground models.)

\section{Sensitivity of Upcoming Interferometers}
\label{detectors}

The MWA, LOFAR and SKA instruments are in various stages of design
planning.\footnote{PAST is furthest along in construction, but it is
not included in our analysis because detailed specifications are not
publically available.  PAST's collecting area is comparable to that of
MWA.}  In our calculations, we try to be faithful to the tentative
design specifications for each facility and to make reasonable
assumptions regarding features of each array that have not been
publicly specified.  Table \ref{table1} lists most of the parameters
for these arrays that we use for our sensitivity calculation.  Unless
otherwise stated, the parameters we adopt come from \citet{bowman05}
for MWA, \citet{deVos04} and www.lofar.org for LOFAR, and
\citet{carilli04} for SKA.

\subsection{Interferometers}
\label{detectors2}
LOFAR will have $77$ large ``stations,'' each of which combines the
signal from thousands of dipole antennae to form a beam of $ \approx
10$ deg$^2$.  Each station is also able to simultaneously image
$N_{p}$ regions in the sky.  We set $N_{p}= 4$ in our calculations,
but this number may be higher.  The signal from these stations is then
correlated to produce an image.  In contrast, MWA will have 500
correlated $4 \, \meter \times 4 \,\meter$ antenna panels, each with
16 dipoles.  This amounts to a total collecting of $7000 ~\meter^2$ at
$z = 8$, or $15\%$ of the collecting area in the core of LOFAR.  While
correlating such a large number of panels is computationally
challenging, this design gives MWA a larger field of view (FOV)
than LOFAR ($800$ deg$^2$), which is an advantage for a statistical survey.

The properties of SKA have not yet been finalized, and it is quite
possible that the EOR science driver for SKA may form a distinct array
from the other, higher-frequency drivers.  In addition, the successes
of MWA and LOFAR will likely influence the final design of SKA.  The
collecting area for SKA is projected to be roughly $100$ times larger
than that of MWA.  There are currently several competing designs for
SKA's antennae.  At one extreme, SKA will have roughly 5000 smaller
antennae (like a much larger MWA).  At the other extreme, it will have
fewer than $100$ large antennae, each of which can simultaneously
image several regions of the sky.  For our calculation, we use the
former extreme case, which makes it easier to have shorter baselines
and to smoothly sample points in the $u$-$v$ plane,---both of which are
important considerations for EOR interferometers.  We assume that the
collecting area for SKA scales as $\lambda^2$, like a simple dipole,
and is equal to $6\times 10^5 \, \meter^2$ within the inner 6 km of
the array for $\lambda = 21 \, (1 + 8)$ cm.  This scaling is somewhat
unrealistic, and the scaling of the collecting area will also depend
on the spacing of the individual dipoles because the antennae will
inevitably shadow each other at the longer wavelengths.\footnote{This
assumes that the low frequency part of SKA consists of dipoles.}

The exact antenna distribution has not been decided for any of these
instruments.  For all three interferometers, we assume that the
distribution of baselines is a smooth function.\footnote{For LOFAR,
which has far fewer antennae units than the other arrays, this
assumption of continuity is fairly crude.}  The distribution of
baselines in an array can substantially impact the sensitivity to the
EOR signal.  For MWA, we calculate the sensitivities for an $r^{-2}$
antenna density profile \citep{bowman05}.  Specifically, this
distribution has a core with a physical covering fraction close to
unity out to $20 \, \meter$ before an $r^{-2}$ falloff and a sharp
cutoff at $750\, \meter$.  The baselines are not as concentrated for
the other two arrays.  LOFAR will have an inner core within 1 km that
has 25\% of its antennae and an outer core with radius equal to 6 km
with another 25\% of its antennae.  For SKA we take (20\%, 30\%, 5\%)
of the antennae within (1, 6, 12 km).  For SKA(LOFAR) we ignore the
antennae outside $12 (6)$ km for our calculations.  For
simplification, we also assume that the density of the antennae is
constant within each outer annulus for LOFAR and for SKA.  However, we
choose the inner 1 km region of both arrays to have a similar $r^{-2}$
distribution to MWA, except with a wider core prior to the $r^{-2}$
falloff in the differential covering fraction.  The lower limit on the
baseline length is approximately $4 \, \meter$ for MWA and $100 \,
\meter$ for LOFAR, and we set this to be $10 \, \meter$ for SKA,
which is approximately the physical diameter of the antennae panels.

For these three arrays, the system temperature is dominated by the sky
temperature.  In our calculations, we set $T_{\rm sys} = T_{\rm sky} =
250 \, K$ at $z= 6$, $T_{\rm sys} = 440 \, K$ at $z = 8$ and $T_{\rm
sys} = 1000 \, K$ at $z = 12$ \citep{bowman05}, and we set $B = 6$ MHz
bandwidth, which translates to a conformal distance of $100 \, \Mpc$
at $z= 8$.  In Appendix \ref{evolution}, we discuss how the choice of
bandwidth can affect observations.  For the sensitivity calculations
in this section, we chose observations that minimized the thermal noise,
-- which is the dominant source of noise on most scales, -- by restricting the
observation for each array to a single FOV.  Finally, we set $\Delta
\nu = 0.01 ~ \MHz$ for all of the arrays.  While these arrays will
have even better resolution than this, improved frequency resolution
does not affect our results.

\begin{deluxetable*}{ c c c c c c c}
\tablecaption{The parameters that we adopt for MWA, LOFAR, and SKA.\tablenotemark{a} }
\tablehead{\colhead{Array} & \colhead{$N_{\rm ant}$} &
\colhead{$N_{\rm ant} \, A_e$ ($\meter^2$)} & \colhead{FOV} &
\colhead{$S_{\rm cut}$\tablenotemark{b}} & \colhead{min. base-} &
\colhead{Cost}\\ & & \colhead{$z= 6/8/12$} & \colhead{($\deg^2$ at
z=8)} & \colhead{($\mu$Jy)} & \colhead{line (m)} & \colhead{($10^6 ~
\$ $)}} \startdata
MWA & 500 & 4500/7000/9000 & $\pi \, 16^2$ & 180 & 4 & $\sim 10$ \\
LOFAR & 64 & $(3.5/4.2/7.2) \times 10^4$ & $4 \times \pi \, 2.0^2$ &
30 & 100 & $\sim 100$\\ SKA & 5000 &
$(3.6/6.0/12.5) \times 10^5$ & $\pi \, 5.6^2$ & 2 & 10 &$\sim
1000$ \\ \enddata \tablenotetext{a}{These are the parameters for a
central region of LOFAR and SKA and not the full array.  We optimize the design for SKA for observations of the EOR, while keeping the current gross specifications for this array.\\}
\tablenotetext{b}{Values for $10^3$ hr of observation with $B = 6 ~ \MHz$ at $150$ MHz.}
\label{table1}
\end{deluxetable*}

\subsection{Results}
\label{sens_results} 

\begin{figure}
\epsfig{file=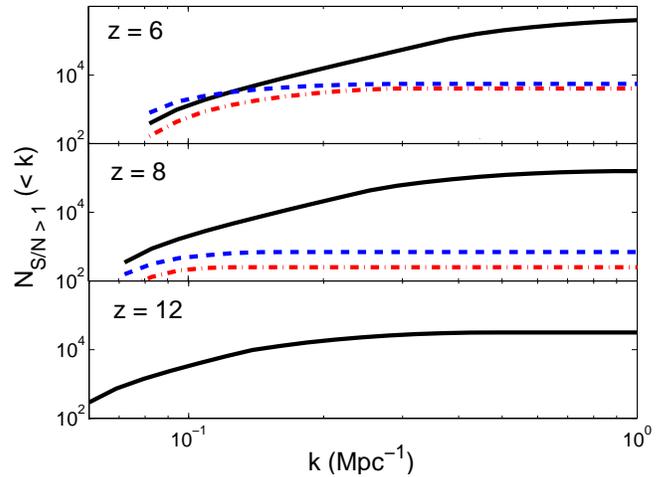, width=8.7cm}
\caption{Integrated number of Fourier pixels $N_{S/N > 1}(k)$ with $k'
< k$ that have a ratio of the RMS signal to the RMS detector noise
that is greater than unity for a $1000$ hr observation with $B = 6 ~
\MHz$.  We use the specifications given in Table \ref{table1} and in
\S \ref{detectors} for MWA ({\it dashed curve}), LOFAR ({\it
dot-dashed curve}), and SKA ({\it solid curve}).  These curves do not
include pixels with $k < 2 \pi/ y$, since foregrounds will contaminate
these pixels substantially.  The 21 cm signal for this calculation is
from a fully neutral medium in which $T_s \gg T_{\rm CMB}$.  If the
universe is partially ionized at $z = 8$, the signal can be both
larger and smaller than this (see Fig. \ref{fig:errors}).  By a redshift of
$12$, only SKA will have any high S/N pixels. }
\label{fig:image}
\end{figure}

\begin{figure}
\epsfig{file=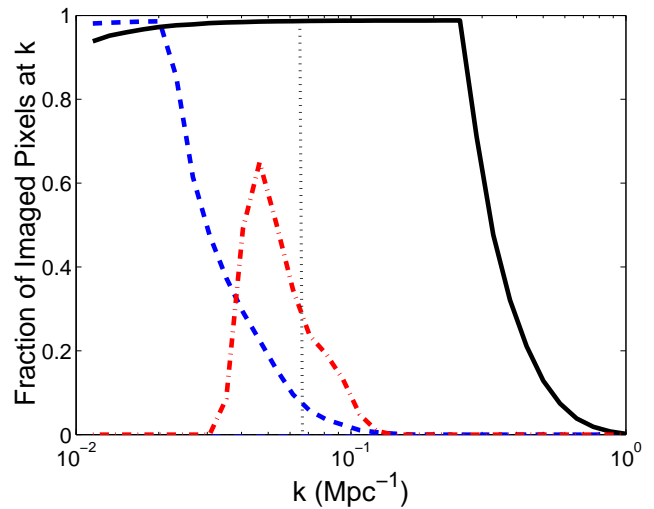, width=8.7cm}
\caption{Fraction of Fourier pixels for MWA ({\it dashed curve}),
LOFAR ({\it dot-dashed curve}), and SKA ({\it solid curve}) that are
``imaged''--- that is, they have a ratio of RMS signal (assuming a neutral
universe) to RMS detector noise that is greater than unity---after $1000$ hr
of observation in a Fourier shell of radius $k$.  The hatched
vertical line marks the depth of this 6 MHz observation at $z = 8$.
Scales to the left of this should be wiped out by foregrounds.  LOFAR
can image a substantially higher fraction of pixels at the relevant
$k$ than MWA, and SKA can image nearly all of its pixels up to $k = 0.3
\, \Mpc^{-1}$. }
\label{fig:image2}
\end{figure}

\begin{figure}
\epsfig{file=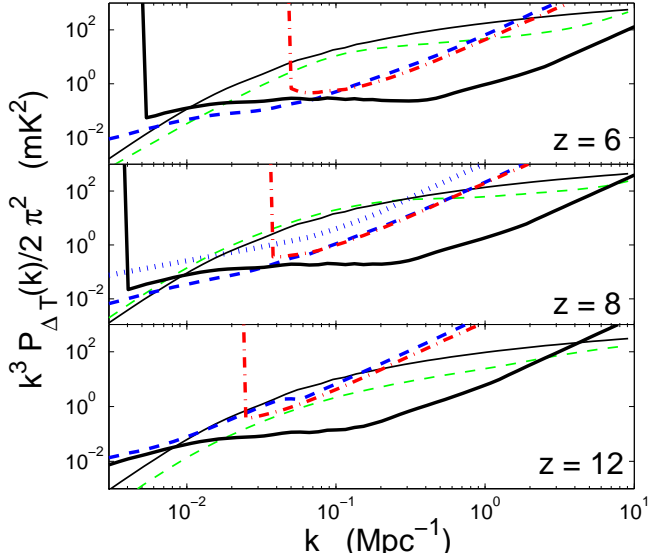, width=8.7cm} 
\caption{Detector noise plus sample variance errors
for a $1000$ hr observation on a single field in the sky, assuming
perfect foreground removal, for MWA ({\it thick dashed curve}), LOFAR
({\it thick dot-dashed curve}), and SKA ({\it thick solid curve}), using the
specifications given in Table \ref{table1} and in \S \ref{detectors}
and for bin sizes of $\Delta k = 0.5 \, k$.  These errors are for
the spherically averaged signal (see text for the discussion of this
point).  The hatched line in the middle panel represents MWA with a flat
distribution of antennae rather than the fiducial $r^{-2}$
distribution.  The detector noise dominates over sample variance for
these sensitivity curves on almost all scales.  The thin solid
curve represents the spherically averaged signal for $\bar{x}_i \ll 1$ and
$T_s \gg T_{\rm CMB}$.  We use this curve to calculate the sample
variance error.  For comparison, the thin dashed curves show the
signal from the FZH04 model when $\bar{x}_i$ is equal to $0.20$,
$0.55$ and $0.75$ for $z = 12, 8$ and $6$, respectively. }
\label{fig:errors}
\end{figure}

Figures \ref{fig:image}, \ref{fig:image2} and \ref{fig:errors} show
the pixel imaging capability and the statistical error in $k^3 \,
P_{\Delta T}(k)/2 \pi^2$, for MWA ({\it dashed curves}), LOFAR ({\it
dash-dotted curves}), and SKA ({\it solid curves}).  For these
figures, we use the parameters given in \S \ref{detectors2} and assume
$1000$ hr of observation over a $6$ MHz band and that the signal comes
from the Universe when $\bar{x}_i = 0$ and $T_s \gg T_{\rm CMB}$.  For
different ionization fractions, the signal can be both larger and
smaller than the assumed signal.  This is illustrated in Figure
\ref{fig:errors}, where the thin solid curves represent the fiducial signal
and the thin dashed curves represent the signal in the FZH04 model for
$\bar{x}_i = 0.2,~ 0.55,$ and $0.75$ for $z = 12, 8$, and $6$,
respectively.

Figure \ref{fig:image} plots the cumulative number of Fourier pixels
for wavenumbers less than $k$ that have ratios of the RMS signal to
the RMS detector noise that are greater than unity. In this plot, we
do not include $k < 2 \pi/y$ in the summation because, as we will
show, the foreground removal procedure makes it unlikely that we can
detect the cosmological signal for values of $k$ smaller than the
depth of the survey. Because MWA has a large FOV and is able to
measure shorter baselines than the other interferometers, it
``images'' a number of Fourier pixels comparable to the number from
LOFAR, despite having less collecting area.  The sensitivity of these
interferometers inevitably declines with redshift, with the detector
noise in a pixel scaling roughly as $T_{\rm sky} \,
\sqrt{\lambda^2/A_e} \sim (1 + z)^{2.6}$, assuming that $A_e \propto
\lambda^2$ and that $T_{\rm sky} \propto \lambda^{2.6}$.  Both LOFAR
and MWA will have fewer than 1000 high signal-to-noise ratio (S/N) pixels at
redshifts 8 and almost no pixels at higher redshifts.  An SKA-class
experiment will be required to image modes with $k > 0.1 ~ \Mpc^{-1}$
or $z \geq 10$.

Observations of high redshift 21 cm emission are promising for
cosmology in part because of the much larger number of Fourier modes
that these observations can probe compared to other cosmological
probes.  These experiments can potentially probe scales larger than
the Jeans length at all times during which the universe is neutral.
The CMB, on the other hand, can only probe primordial fluctuations up
to the Silk damping scale (${\it l}_{\rm Silk} \approx 4000$) from a
single angular power spectrum.  Currently, CMB experiments can image
${\it l}_{\rm max}^2 < {\it l}_{\rm Silk}^2$ independent modes.  The
Wilkinson Microwave Anisotropy Probe (WMAP) is cosmic variance-limited
for modes smaller than ${\it l}_{\rm max} \approx 400$, and Planck is
cosmic variance-limited by ${\it l}_{\rm max} \approx 3000$.  The
number of modes that can be imaged by SKA in a $1000$ hr observation
at $z = 6$ in a $6 \, \MHz$ band is larger than the number of imaged
modes for WMAP and significantly less than this number for Planck.  A
longer observation or a larger bandwidth will increase the number of
modes that these interferometers can observe.
	
The reason the $S/N$ is generally smaller for 21 cm measurements than
it is for measurements of the CMB is in part due to the bandwidth of
these observations.  Both cosmological probes are looking for
fluctuations that are of order $10^{-5}$ times that of the sky temperature.
Unfortunately, the number of independent samples of the sky
temperature is proportional to the bandwidth, and CMB experiments have
$\sim 1000$ times larger bandwidth since they observe at $\nu \approx
100 \,{\rm GHz}$.  Therefore, CMB experiments can beat down their
uncertainty in $T_{\rm sky}$ by an additional factor of $(1000)^{1/2}$
for an observation of the same duration.

 Figure \ref{fig:image2} plots the {\it fraction} of pixels for a
 given value of $k$ with a ratio of RMS signal to RMS noise greater
 than unity for MWA, LOFAR, and SKA.  The vertical hatched line indicates $2
 \pi/y$, and it is likely that foregrounds can be cleaned well enough
 only at scales rightward of this line.  A larger fraction of LOFAR's pixels
 than MWA's pixels will have high S/N. Because MWA has a larger FOV,
 it still can detect a comparable number of high S/N pixels (see
 Fig. \ref{fig:image}).\footnote{LOFAR should, however, have more
 success imaging a single quasar on the sky than MWA because it can
 image a larger fraction of its pixels.  Since MWA is extremely cored,
 this will make its beam much more coarse than the other experiments.}
 SKA will have high S/N detections in almost all of its pixels up to
 $k \sim 0.3 ~ \Mpc^{-1}$.  Figure \ref{fig:image2} illustrates that
 if foregrounds contaminate more large wavelength modes than is
 assumed, MWA and LOFAR can have substantially fewer high S/N pixels.
 Alternatively, a larger bandwidth will result in more high S/N
 pixels.

Figure \ref{fig:errors} compares the interferometers' ability to
statistically constrain $P_{\Delta T}(k)$, ignoring the effect that
foregrounds have on the sensitivity.  Even though $P_{\Delta T}(k)$ is
not spherically symmetric, we spherically average $P_{\Delta T}$, as
well as the errors, for the purpose of this plot.  Because of this
averaging, these interferometers will be slightly more sensitive to
some modes than this plot implies.  At z = 6, the trend is as
expected: SKA is more sensitive than LOFAR and LOFAR is more sensitive
than MWA.  Still, LOFAR's gains over MWA are not proportional to the
square of the collecting area, as we might naively expect.  At higher
redshifts, LOFAR and MWA are comparably sensitive on most scales.  We
also plot the sensitivity of MWA at $z = 8$ for a flat distribution of
antennae rather than the fiducial $r^{-2}$ distribution of antennae.
In this case, MWA is substantially less sensitive at {\it all} scales.
This contrasts with angular power spectrum measurements, where a flat
distribution of antennae is always more sensitive at larger $k$ than a
tapered distribution.

If all the arrays had the same normalized distribution of baselines,
and if the error on the measurement of $P_{\Delta T}(\bfk)$ in a
Fourier pixel scales inversely with the square of the differential
covering fraction for that $\bfk$, as it does for the angular power
spectrum, LOFAR should be many times more sensitive than MWA to a
given pixel, -- at least in the case where detector noise is the
dominant source of noise.  Because MWA observes many more independent
Fourier cells owing to a larger survey volume, for statistical detections, MWA
should fare better even with the same distribution of baselines.
However, the normalized distribution of baselines is not the same for
these arrays.  All these interferometers have a similar covering
fraction in the very center, since the maximum covering fraction is
unity.  Since MWA stacks all its antennae in the core, it does not
have any baselines that probe $k_{\perp} > 0.5 ~\Mpc^{-1}$ at $z =8$.
Unlike MWA, only a quarter of LOFAR's antennae are in its core
region, leading to a smaller fraction of its baselines that can
observe modes with $k_{\perp} < 0.5 ~ \Mpc^{-1}$.  Another reason that
LOFAR is not many times more sensitive than MWA is because LOFAR's
minimum baseline of $100 ~\meter$ does not allow it to detect modes
with $k_{\perp} \la 0.03 ~ \Mpc^{-1}$ (Fig. \ref{fig:errors}). These
modes happen to be those to which MWA is most sensitive.

\begin{figure}
\epsfig{file=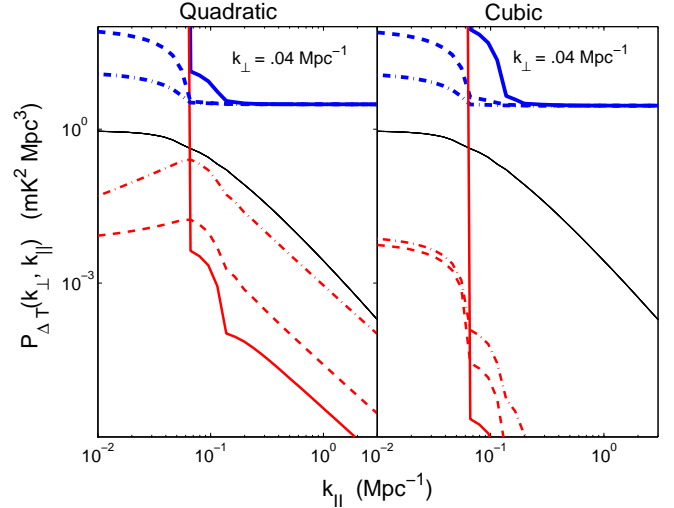, width=8.7cm}
\caption{Foreground removal for MWA for a $1000$ hr observation with
$B = 6~\MHz$. The thin solid curve shows $P_{\Delta T} (\bfk_\perp,
k_{||})$ plotted as a function of $k_{||}$ for a neutral universe.
The thick upper curves show the sensitivity to the signal $\delta
P_{\Delta T} (\bfk_\perp, k_{||})$ if we remove the foregrounds using
the method outlined in \S \ref{foregrounds} in a band centered on the
$6$ MHz cosmological signal processing window of $6, 12,$ or $24 ~
\MHz$ ({\it solid, dashed or dot-dashed curves, respectively}).  Note
that the sensitivity curves are above the curves for signal because
this is the sensitivity to a single Fourier pixel ($N_c = 1$).  The
lower curves show the function $Q_{\bfk_{\perp}}$, -- the residual
foreground level, -- for each of the three processing window bandwidths
(see \S \ref{foregrounds}). The foreground power in each panel is
assumed to be a 1-$\sigma$ fluctuation of the model discussed in \S
\ref{foregrounds}.  The left panel shows a fit with a quadratic
polynomial (${\rm n}=2$) and the right shows a fit with a cubic polynomial
(${\rm n}=3$).  While increasing ${\rm n}$ will always remove more of
the foregrounds: it will also remove more of the cosmological signal
as well.  This can be seen in this plot by the reduced sensitivity to
the signal at large scales. (In the absence of foregrounds and
foreground cleaning, the sensitivity curves would be flat in this
plot.) }
\label{fig:foregrounds1}
\end{figure}

\begin{figure}
\epsfig{file=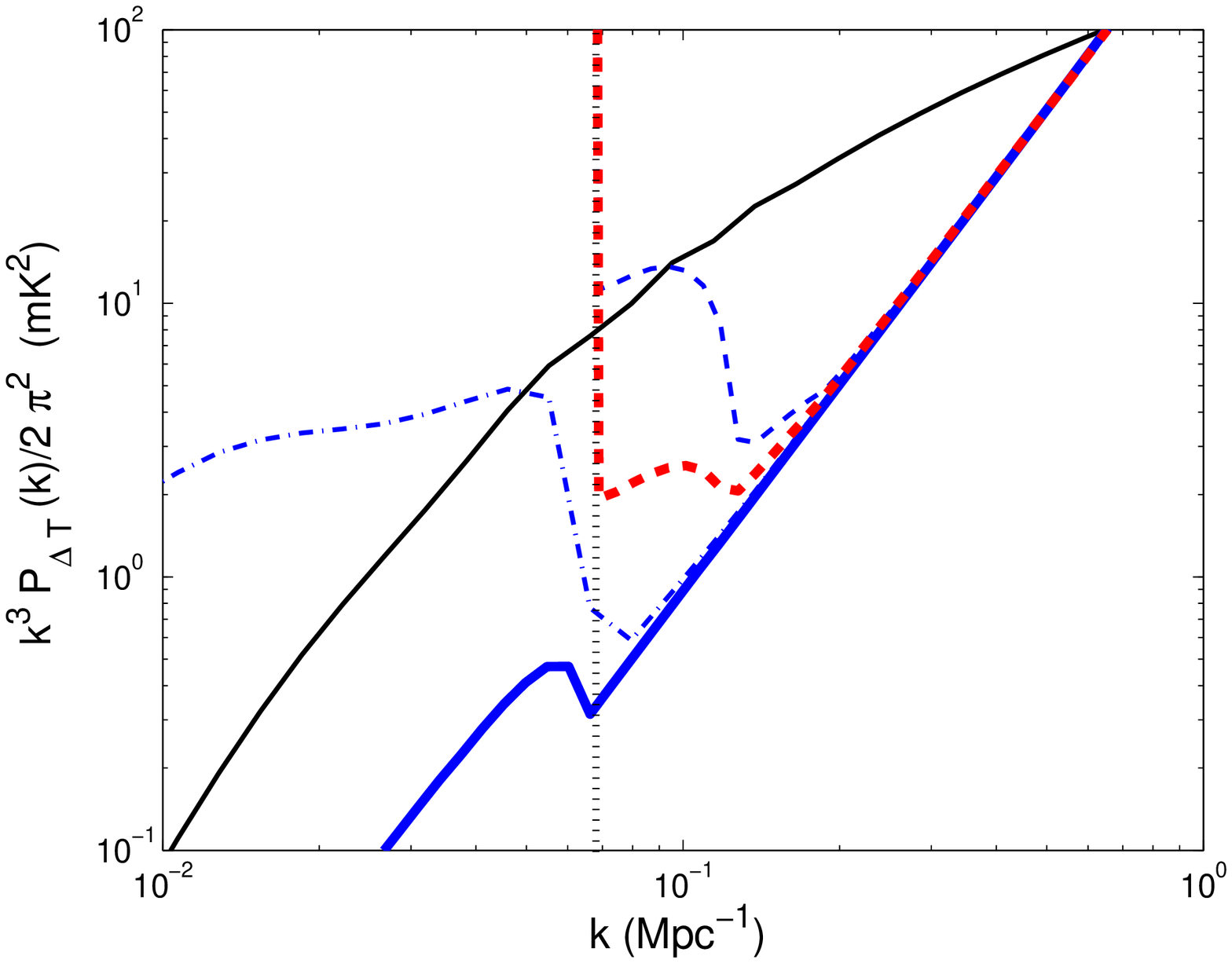, width=8.7cm}
\caption{Dimensionless 21 cm power spectrum for a neutral universe
measured in a $6~ \MHz$ band ({\it thin solid curves}) and the
1-$\sigma$ error for a $1000$ hr observation by MWA in the absence of
foregrounds ({\it thick solid curves}), after foreground subtraction
with a cubic polynomial at all $k_{\perp}$ in a $6 ~\MHz$ window ({\it
dashed curve}) and in a $24 ~\MHz$ window ({\it dot-dashed curve}).  The
thick dashed curves is for a quadratic polynomial from a $6 ~\MHz$
band.  The vertical hatched line denotes the scale of the $6~ \MHz$ box.
Foreground cleaning reduces the sensitivity at large scales.  This
plot illustrates that it makes a big difference how foregrounds are
removed from the signal: experiments will want to remove the
foregrounds over a fairly large bandwidth and with as low order a
polynomial as possible.}
\label{fig:foregrounds2}
\end{figure}

Figures \ref{fig:foregrounds1} and \ref{fig:foregrounds2} illustrate
how foregrounds affect the sensitivity of MWA to the power spectrum
for a fixed $\bfk_{\perp}$ and to statistical detections of the power
spectrum.  The foregrounds are first fitted in the frequency direction
for each $\bfk_{\perp}$.  As we will see, the foreground preprocessing
removes significant power from the signal on large scales, reducing
the sensitivity to the signal at such scales.  In Figure
\ref{fig:foregrounds1}, we assume a 1-$\sigma$ angular fluctuation for
the foreground model outlined in \S \ref{foregrounds}.  We then
subtract a quadratic ({\it left panel}) or cubic polynomial ({\it
right panel}) from the foregrounds over a frequency interval of $6$,
$12$, and $24$ MHz.  For these figures, the $6$ MHz band from which we
extract the 21 cm signal is centered in the larger frequency intervals
in which we remove the foregrounds.  The placement of this band does
not affect the results substantially.  We find that for all the
bandwidths, a quadratic or a cubic polynomial is able to remove the
residual foregrounds $Q_{\kvec_\perp}$, defined in equation
(\ref{eq:cond}), substantially below the signal ({\it thin solid
line}).  The solid, dashed and dash-dotted curves at the bottom of
Figure \ref{fig:foregrounds1} indicate $Q_{\kvec_\perp}$ for
foreground removal in a $6$, $12$ and $24$ MHz band.  The cubic
polynomial is able to remove the foregrounds well below the signal for
all cases, but for $B \la 12~ \MHz$ (or at larger $\bfk_{\perp}$ than
shown, where the foreground contamination is smaller) the quadratic
polynomial is sufficient.  This conclusion holds for the other
interferometers as well.

Foreground subtraction removes power from the cosmological signal on
smaller scales as we decrease the bandwidth over which we remove the
foreground (or as we increase the order of the polynomial).  Our
analysis accounts for this by effectively dividing the sensitivity
curves by the filter function that describes how power is removed from
the cosmological signal as a function of $k_{||}$ (Fig. \ref{fig:foregrounds1}, {\it upper thick curves}).  These sensitivity curves would
be flat in the absence of foregrounds and foreground cleaning.  The
foreground cleaning causes these curves to be less sensitive to the
signal at large scales.  The errors on the power spectrum are
substantially smaller for the second generation of interferometers,
but the foreground removal has a qualitatively similar effect on these
interferometers' sensitivity curves.

We can also combine the signal along different $\bfk_{\perp}$ in
Fourier space to constrain $P_{\Delta T}(\bfk)$
(eqn. \ref{eq:forstat}).  Figure \ref{fig:foregrounds2} shows the
statistical error in the spherically averaged 21 cm power spectrum for
MWA after foregrounds are cleaned.  The thick solid line represents
the error just from detector noise.  At around the scale corresponding
to the depth of the box, too much power from the cosmological signal
is removed due to foreground cleaning for MWA to be sensitive.  As
before, this effect is minimized by fitting to a larger bandwidth.
The dashed and dash-dotted curves indicate the errors if we remove the
foregrounds with a cubic polynomial (${\rm n} =3$) in a $6$ and
$24~\MHz$ band, respectively.  The thick dashed curve shows the errors
for a quadratic polynomial (${\rm n}=2$) with $B= 6 ~\MHz$: ${\rm
n}=2$ removes substantially less power than ${\rm n} = 3$ given the
same bandwidth.  Similar conclusions hold for other EOR
interferometers.

Despite the simple model we employ for foregrounds, we expect our
conclusions pertaining to foreground removal to be fairly robust.  Our
technique should be able to remove the foregrounds from a mode with
$k_{||} \gg 2 \pi/y$ as long as $P^{f}(\bfk) \ll P^{21}(\bfk)$, where
$P^{f}$ is the foreground 3-D power spectrum.  The foregrounds are
expected to be in this limit for most relevant $\bfk$.  The process of
removing the foregrounds from the signal will inevitably remove the
signal for $k_{||} \la 2 \pi/y$.

This is not to say that removing foregrounds from 21 cm maps is
trivial.  Our analysis neglected several complications that the real
observations must deal with.  Since the observed wavelength increases
with redshift, over the depth of the survey a mode with a set
value of $\bfk_\perp$ will be measured by different baselines.  In our
analysis, we ignored this effect.  As long as the distribution of
baselines is fairly smooth, we expect that this will have a minor
effect on foreground removal.  Other foregrounds that are beyond the
scope of this paper include residuals owing to imperfect point source
subtraction, radio frequency interference contaminating frequency
intervals within the observation band (this may be a substantial
challenge for LOFAR, which is in a radio loud environment), and the
residuals that arise owing to the imperfect modeling of atmospheric
distortions (modeling the atmosphere may be a significant challenge
for MWA due to its large FOV).

\section{Cosmology from the 21 cm Power Spectrum}
\label{cosmology}

Observations of high redshift 21 cm emission are capable of measuring
$P_{\delta \delta}$ on smaller scales than current CMB experiments.
Our calculations show that SKA can sensitively probe comoving
megaparsec scales, which are also smaller than scales observed by
galaxy surveys and comparable to scales probed with the Ly$\alpha$
forest.  The sensitivity to smaller scales than the CMB may allow 21
cm observations to break degeneracies among cosmological parameters
that are present in CMB constraints.

In this section, we utilize the sensitivity calculation described in
\S \ref{noise} and \S \ref{detectors} to estimate how well upcoming
EOR interferometers can constrain cosmological parameters from
$P_{\Delta T}$, and, in particular, whether these constraints will be
competitive with CMB observations.  We divide our discussion into
three cases: (1) If density fluctuations dominate the signal (\S
\ref{mu4}).  This can happen if $\bar{x}_i \ll 1$ and $T_{\rm CMB} \ll
T_{s}$ or if X-rays are responsible for the reionization of the
Universe. (2) When the bubbles contaminate the $P_{\mu^0}$ and
$P_{\mu^2}$ terms such that only $P_{\mu^4}$ and $P_{\mu^6}$, -- which
arises from the AP effect, -- are pristine enough to measure
cosmological parameters (\S \ref{ap}).  (3) On large scales at which
neutral fraction fluctuations are important, but at which these
fluctuations trace the density fluctuations (\S \ref{largescales}).
Note that the analysis in this section does {\emph not} assume any
model for reionization.

It came to our attention that \citet{bowman05b} was performing a
similar analysis for MWA and MWA5000.  This paper, submitted
concurrently with ours, pertains to when the signal is in the regime
we discuss in \S\ref{mu4}.  There are a few differences between our
two approaches.  In addition to cosmological parameters,
\citet{bowman05b} fits to parameters for the foreground residuals as
well as other observational parameters. \citet{bowman05b} also assumes
a spherically symmetric $P_{\Delta T}$.  Deviations from spherical
symmetry enhance cosmological parameter constraints.  Another
important difference is that our analysis combines 21 cm observations
with current and future CMB experiments.  This can break parameter
degeneracies present in these separate cosmological probes, and is
important for assessing the true value of 21 cm observations for cosmology.

\subsection{Density Fluctuations Dominate}
\label{mu4}

We first concentrate on the signal in the case where the density
fluctuations dominate over the spin temperature and neutral fraction
fluctuations.  This is the case in which 21 cm observations will be
most sensitive to cosmological parameters.  If reionization occurred
at $z \approx 6.5$, as the Sloan quasars suggest \citep{becker01,
fan02}, it is not altogether unlikely that upcoming interferometers
will observe the signal in this regime.  Models in which $\bar{x}_i =
\zeta \, f_{\rm coll}$ have a significant period in which $\bar{x}_i
\ll 1$.  During this period, fluctuations in $x_H$ are unimportant.
In addition, it is expected that at higher redshifts than those considered
here, X-ray photons and shocks heat the gas in the IGM to well above
the CMB temperature \citep{venkatesan01,
chen04}. \citet{ciardi03-21cm} argue that around $z = 20$ the first
stars will produce a large enough background in Ly$\alpha$ such
that $T_s$ will be coupled to the kinetic temperature of the gas
through the Wouthuysen-Field effect.  If this is true, spin
temperature fluctuations will be subdominant at the redshifts we
consider.

Tables \ref{table3} and \ref{table4} quantify how the 21 cm signal can
constrain some of the most interesting cosmological parameters:
$\tau$, $\Omega_w, w$, $\Omega_m \,h^2$, $\Omega_b \,h^2$, $n_s$,
$\delta_H$, $\alpha_s$ and $\Omega_\nu$.  The tilt $n_s$ we define to
be the power law index of the primordial power spectrum at $k = 0.05
\, \Mpc^{-1}$ and $\alpha_s = dn_s(k)/d\log(k)$.  The parameter
$\delta_H$ is roughly the size of density fluctuations at the present
day horizon scale (defined here such that the primordial power
spectrum today $P(k) = 2 \pi^2 \, \delta_H^2 \, k^{n_s(k)}/[70\, {\rm
km \, s^{-1} \, Mpc^{-1}}]^{3 +n_s(k)}$).  To construct the linear
power spectrum used in this analysis, we employ the transfer function
from the code CAMB.\footnote{Available at http://camb.info.}  To get
confidence intervals, we use the Fisher matrix formalism
\citep{tegmark97b}.

In the long term, MWA plans to increase the number of antenna panels
from 500 to 5000.  This array, MWA5000, will have a comparable
collecting area to LOFAR and $10\%$ of the collecting area of SKA.  To
model MWA5000, we use an $r^{-2}$ distribution of antennae out to 1
km, similar to MWA, but with a larger flat core than MWA that extends
out to $80 \, \meter$ rather than $20 \, \meter$.  We also include
MWA50K, which is another 10 times larger than MWA5000, but again built
in the same mold as MWA.  Correlating 50,000 antennae will be a
significant computational challenge.

 In Table \ref{table3}, we calculate the 1-$\sigma$ errors on
cosmological parameters for observations at $z = 8$  in the case in which
$\bar{x}_i \ll 1$ and $T_s \gg T_{\rm CMB}$, such that all
the $\mu$ terms trace the density power spectrum.  Unless otherwise
noted, we perform the calculations in this section for an observation
of $2000$ hr on two locations in the sky, or roughly $2$ productive
years.\footnote{Interferometric observations have never been integrated for
such long periods on a single field.  It is uncertain whether such
observations are even possible, and this will depend heavily on how
well we can deal with various systematics in these systems.}
Observations should generally be chosen to minimize the number of
patches on the sky and maximize the duration for which each patch is
observed because detector noise dominates the uncertainty at most
scales. For the second generation of EOR interferometers, this is not
necessarily the case, and sometimes parameter estimates are improved
by choosing a different observing strategy.

Future observations have the potential to improve many of the current
constraints on cosmological parameters.  Two years of observation with
MWA and LOFAR have trouble constraining a five parameter cosmology:
$\Omega_\Lambda$, $\Omega_m\, h^2$, $\Omega_b \, h^2$, $n_s$, and the
normalization parameter $\bar{x}_H \,\delta_H$ (see Table \ref{table3}, note
`b').  However, when combined with current CMB observations (WMAP,
Boomerang, ACBAR [Arcminute Cosmology Bolometer Array Receiver], and
CBI [Cosmic Background Imager]), both measurements by MWA and LOFAR
are able to improve measurements of $\Omega_\Lambda, \Omega_m \, h^2,
\Omega_{\nu}, n_s,$ and $\alpha_s$.  MWA and LOFAR are not able to
significantly improve the constraints from Planck.  Unfortunately,
$\bar{x}_H$ is not well constrained by measurements by MWA or by LOFAR
when they are combined with current CMB observations.  This is because
the current uncertainty in cosmological parameters leads to
substantial uncertainty in the amplitude of $P_{\delta_L \delta_L}$ at
relevant scales.  Planck will be able to refine the measurement of
these parameters and the first generation of 21 cm experiments plus
Planck will place tighter constraints on $\bar{x}_H$.

\begin{figure}
\epsfig{file=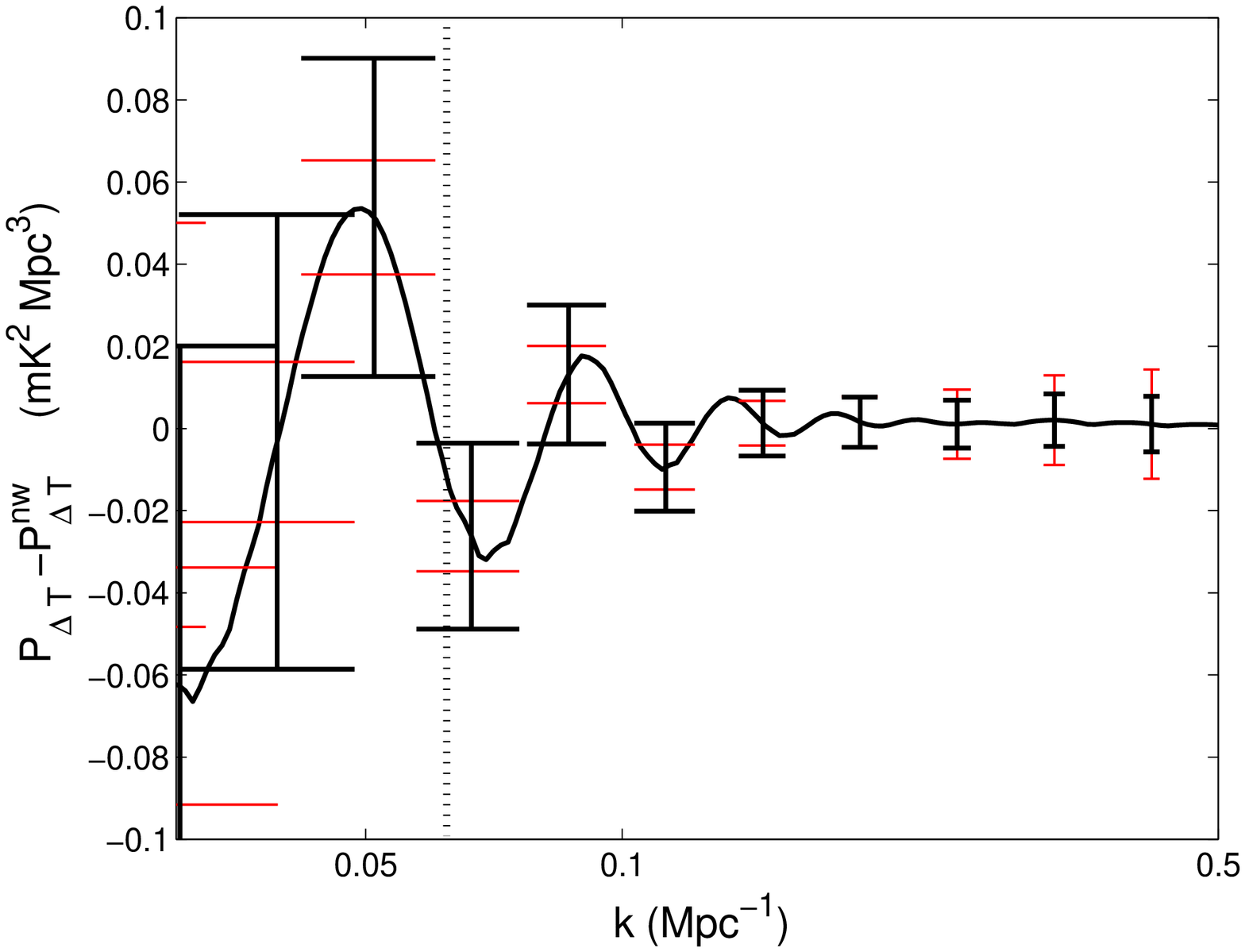, width=8.7cm}
\caption{Cosmic variance plus detector noise errors on the large scale
modes of the spherically averaged power spectrum for SKA ({\it thick error
bars}) and MWA5000 ({\it thin error bars}) after $1000$ hr of observation in a
6 MHz band.  We accentuate the wiggles in $P_{\Delta T}(k)$ by
subtracting a power spectrum that does not have baryonic wiggles
$P^{\rm nw}_{\Delta T}(k)$.  The vertical hatched line indicates the
size of a 6 MHz box, approximately where we expect foregrounds to
swamp the signal.  The second generation of interferometers will be
able to detect the wiggles at a few $\sigma$ significance level.  MWA5000
is more sensitive than SKA to these features because of its larger
FOV.  The signal here is for $\bar{x}_i \ll 1$; the presence of HII
bubbles may enhance these wiggles (\S \ref{largescales}).}
\label{fig:wiggles}
\end{figure}

The second or third generation of 21 cm observations will be
substantially more sensitive to the cosmology.  By themselves,
MWA5000, SKA and MWA50K can constrain a seven parameter cosmology that
involves $\alpha_s$ and $\Omega_{\nu}$ in addition to the other five
parameters we used for LOFAR and MWA (Table \ref{table3}).
Surprisingly, MWA5000 is comparably sensitive to SKA despite having
$10$ times less collecting area.  This is because SKA is not as
centrally concentrated with only 20\% of its antennae in the 1 km core
while MWA5000 has 100\%.  Also, MWA5000 has a larger FOV than SKA,
which results in smaller errors on large scales, scales at which these
arrays are sample variance limited.  Large scales probe the baryonic
wiggles and therefore can provide substantial constraining power
(Fig. \ref{fig:wiggles}).  If we alter the observation for SKA to
decrease the sample variance, -- having it observe ten locations on the
sky each for $400$ hr, -- then SKA's sensitivity is improved (see
entries with asterisks in Table \ref{table3}).

 In combination with Planck, MWA5000 and SKA can improve constraints
on $\Omega_w$, $\Omega_m \,h^2$, $n_s, \alpha_s$ and $\Omega_\nu$, and
MWA50K can do even better (Table \ref{table3}).  Because these
observations probe smaller scales than the CMB, the parameters that
affect the small scale behavior, namely $n_s, \alpha_s$ and
$\Omega_\nu$, show the most substantial improvement.  In addition, as
one changes the cosmological parameters in the conversion from $\uvec$
to $\bfk$, this distorts the measured power spectrum in k-space,
providing an additional effect that can be used to constrain
parameters.  This is illustrated in Table \ref{table3}: the
constraints on SKA denoted with superscript `c' are if we do not vary
cosmological parameters in the conversion from $\uvec$ to $\bfk$.  The
uncertainty on $\Omega_{\Lambda}$, $\Omega_{m}\, h^2$, and
$\Omega_{\nu}$ is substantially larger in this case.

Table \ref{table4} shows the sensitivity to cosmological parameters if
the above scenario occurs at higher redshifts than $z = 8$.  MWA5000
is still sensitive to the signal at $z = 10$, and its sensitivity
falls off at $z = 12$.  Our design for SKA is unrealistically
optimized for all considered redshifts.  Because of this, SKA is more
sensitive than MWA5000 at $z = 12$, but it sensitivity is still
falling due to the increasing sky temperature.

The uncertainty estimates in this section are for
observations with $B = 6 ~\MHz$.  Experiments will be able to
process a much larger bandwidth, and, if we
are fortunate, nature could provide an even larger redshift slice in which
density fluctuations dominate.

 \begin{deluxetable*}{l c c c c c c c c c c}
\tablecolumns{11}
\tabletypesize{\footnotesize}
\tablecaption{Errors on cosmological parameter estimates when density fluctuations dominate the 21cm signal.\tablenotemark{a}}

\tablehead{ & \colhead{$\tau$} & \colhead{$\Omega_w$} & \colhead{$w$} & \colhead{$\Omega_m \,h^2$} & \colhead{$\Omega_b \,h^2$} & \colhead{$n_s$} &  \colhead{$\delta_H \times 10^5$~\tablenotemark{b}} & \colhead{$\alpha_s$} & \colhead{$\Omega_\nu$} & \colhead{$\bar{x}_H$}}
\startdata
& 0.1 & 0.7 & -1.0 & 0.14 & 0.022 & 1.0 & 3.91 & 0.0 & 0.0 & 1.0 \\
\hline
LOFAR & - & 0.07 & - & 0.11 & 0.03 & 0.11 & 5.0 & - & - &- \\
MWA & - & 0.06 & - & 0.09 & 0.02 & 0.09 & 4.2 & - & - &- \\
MWA5000 & - & 0.005 & - & 0.008 & 0.002 & 0.03 & 0.37 & 0.010 & 0.007 & -\\
SKA &  - & 0.005 & - & 0.009 & 0.002 & 0.06 & 0.51 & 0.016 & 0.015 & - \\
SKA\tablenotemark{c} & -& 0.11 & - & 0.042 & 0.003 & 0.07 & 2.0 & 0.017 & 0.08 &- \\
SKA\tablenotemark{*} &  - & 0.004 & - & 0.007 & 0.002 & 0.03 & 0.32 & 0.010 & 0.008 & -  \\
MWA50K\tablenotemark{*} & - & 0.002 & - & 0.004 & 0.001 & 0.01 & 0.17 & 0.004 & 0.002 & - \\
\hline
CCMB & 0.060 & 0.084 & - & 0.017 & 0.0014 & 0.072 & 0.29 & 0.039 & 0.12 & - \nl
CCMB+ LOFAR  & 0.057 & 0.050 & - & 0.010 & 0.0012 & 0.027 & 0.22 & 0.022 & 0.02 & 0.2 \\
CCMB+ MWA  & 0.056  & 0.046 & -  & 0.009 & 0.0011 & 0.021 & 0.22 & 0.022 & 0.02 & 0.2 \\
CCMB+ MWA5000  & 0.048 & 0.005 & - & 0.003 & 0.0009 & 0.013 & 0.18 & 0.005 & 0.004 & 0.06  \\
CCMB + SKA & 0.048 & 0.005 & - & 0.003  & 0.0009 & 0.014 & 0.18 & 0.005 & 0.007 & 0.06\\
\hline
Planck & 0.0050 & 0.029 & 0.09 & 0.0023 & 0.00018 & 0.0047 & 0.026 & 0.008  & 0.010 & - \\
Planck +MWA5000 & 0.0046 & 0.017 & 0.06 & 0.0009 & 0.00012 & 0.0033 & 0.018 & 0.003  & 0.003 &0.03\\
Planck + SKA & 0.0046 & 0.021 & 0.08 & 0.0008 & 0.00012 & 0.0034 & 0.018 & 0.003 & 0.004 & 0.04 \\
Planck + SKA\tablenotemark{*} & 0.0046 & 0.017 & 0.07 & 0.0007 & 0.00012 & 0.0032 & 0.017 & 0.003 & 0.003 & 0.03 \\
Planck + MWA50K\tablenotemark{*} & 0.0045 & 0.007 & 0.03 & 0.0004 & 0.00010 & 0.0029 & 0.016 & 0.002 & 0.001 & 0.01\\
\enddata
\tablenotetext{a}{We assume, unless otherwise noted, observations of
 $2000$ hr on two places in the sky in a $6 ~\MHz$ band which is
 centered at $z = 8$.  Current CMB (CCMB) are the combined results of WMAP,
 Boomerang, ACBAR, and CBI.  In these calculations, we account for
 foregrounds by imposing a sharp cutoff in sensitivity at $k = 2
 \pi/y$, where $y$ is the width of the box, and we avoid fitting to
 scales in the non-linear regime by imposing a small scale cutoff at
 $k = 2 ~ \Mpc^{-1}$.  These calculations are for a flat universe, $1
 = \Omega_m + \Omega_w$, and dashes indicate parameters which are not
 marginalized.\\}
\tablenotetext{b}{From just the 21 cm data, the parameter $\delta_H$
 is completely degenerate with $\bar{x}_H$.  Because of this, for 21
 cm observations alone, the constraints in this column are really for
 the parameter $\bar{x}_H \, \delta_H$.\\}
\tablenotetext{c}{We use the fiducial cosmology in the conversion from $\bfu$ to $\bfk$ such that the angular diameter distance and the depth of the map do not change when we vary parameters to get the above confidence intervals.\\}
\tablenotetext{*}{Observations of $10$ locations on the sky, at $400$
hr each.}
\label{table3}
\end{deluxetable*}

\begin{deluxetable*}{ l c c c c c c c c c}
\tablecaption{Same as Table \ref{table3}, but for higher redshifts.\tablenotemark{a}}
\tabletypesize{\footnotesize}
 \tablehead{ & \colhead{$\tau$} & \colhead{$\Omega_\Lambda$} & \colhead{$\Omega_m \,h^2$} & \colhead{$\Omega_b \,h^2$} & \colhead{$n_s$} &  \colhead{$\delta_H \times 10^5$~\tablenotemark{b}} & \colhead{$\alpha_s$} & \colhead{$\Omega_\nu$} & \colhead{$\bar{x}_H $}}

\startdata
 & 0.1 & 0.7 & 0.14 & 0.022 & 1.0 & 3.91 & 0.0 & 0.0 & 1.0 \\
\hline
MWA5000 ($z = 10$) &- & 0.010 & 0.014 & 0.004 & 0.03 & 0.6 & 0.01 & 0.010 & - \\
SKA ($z =10$)\tablenotemark{*} & - & 0.007 & 0.010 & 0.003 & 0.03 & 0.4 & 0.01 & 0.009 & - \\
MWA5000 ($z = 12$) &- & 0.019 & 0.030 & 0.008 & 0.07 & 1.4 & 0.03 & 0.016 & - \\
SKA ($z =12$)\tablenotemark{*} & - & 0.011 & 0.014 & 0.004 &  0.05 & 0.7 & 0.02 & 0.013 & - \\
\hline
Planck & 0.0049 & 0.011 & 0.0023 & 0.00017 & 0.0047 & 0.03 & 0.007 & 0.010 & - \\
Planck + MWA5000 ($z =10$) & 0.0047 & 0.007 & 0.0013 & 0.00013 & 0.0036 & 0.02 & 0.005 & 0.003 & 0.03 \\
Planck + SKA ($z =10$)\tablenotemark{*} & 0.0046 & 0.006 & 0.0011 &  0.00013 & 0.0035 & 0.02 & 0.004 & 0.003 & 0.03 \\
Planck + MWA5000 ($z =12$) & 0.0049 & 0.009 & 0.0017 & 0.00015 & 0.0040 & 0.02 & 0.007 & 0.004 & 0.04 \\
Planck + SKA ($z =12$)\tablenotemark{*} & 0.0047 & 0.007 & 0.0014 &  0.00014 & 0.0037 & 0.02 & 0.005 & 0.004 & 0.04 \\
\enddata
\tablenotetext{a}{See table note (a) in Table \ref{table3} for the specifications used in these calculations.\\}
\tablenotetext{b}{From just the 21 cm data, the parameter $\delta_H$ is completely degenerate with $\bar{x}_H$.  Because of this, for the 21 cm observations alone, the constraints in this column are really for the parameter  $\bar{x}_H \, \delta_H$.\\}
\tablenotetext{*}{Observations of $10$ locations on the sky, at $400$ hr each.\\}
\label{table4}
\end{deluxetable*}

\subsection{Neutral Fraction Fluctuations are Significant}
\label{ap}

\begin{figure}
\epsfig{file=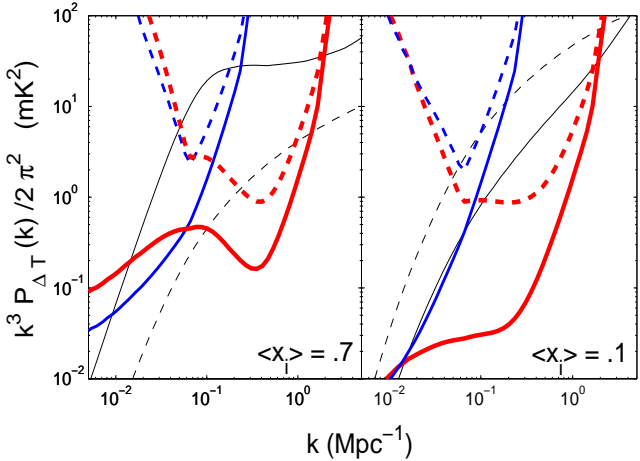, width=8.7cm, height=6.3cm}
\caption{Signal using the FZH04 model ({\it thin curves}) and detector
noise plus cosmic variance errors ({\it all other curves}) at $z=8$
for a $2000$ hr observation combining two different fields of view, or
roughly $2$ yr of observations, for the $P_{\mu^0}$ ({\it solid
curves}) and $P_{\mu^4}$ ({\it dashed curves}) components of the
signal.  In the left panel $\bar{x}_i = 0.7$, and in the right panel
$\bar{x}_i = 0.1$.  We plot the sensitivity curves for MWA ({\it
medium-width curves}) and for SKA ({\it thick curves}) with a binning
width of $\Delta k = 0.5\, k$.  These curves assume that the angular
diameter distance to $z$ and that the value of $H(z)$ are well
constrained such that we can convert between $\bfu$ and $\bfk$.  If
this is not the case, there is also a $\mu^6$ term.}
\label{fig:mudecomp}
\end{figure}

In this section, we investigate whether it is possible to extract
$P_{\mu^4} = \bar{x}_H^2 \, P_{\delta_L \delta_L}$ from $P_{\Delta
T}(\bfk)$ well enough to constrain cosmological parameters when
ionized fraction fluctuations are important.\footnote{The techniques
in this section also apply to periods during which spin temperature
fluctuations are important.}  Figure \ref{fig:mudecomp} shows the $z =
8$ sensitivity curves for MWA and for SKA. The right panel is shown at
the beginning of reionization ($\bar{x}_i = 0.1$), when the density
fluctuations are still the largest source of fluctuations.  In this
case, SKA will be sensitive to $P_{\mu^4}$ over 1-2 decades in $k$.
Conversely, MWA is not sensitive to $P_{\mu^4}$. The left panel shows
the opposite case, when the bubbles dominate ($\bar{x}_i = 0.7$). In
this case, MWA and SKA are both not sensitive to $P_{\mu^4}$.  Both
MWA and SKA can be sensitive to the $P_{\mu^{0}}$ over a range of
scales.  This analysis assumes that the fiducial cosmology is correct
or else we could not measure $P_{\mu^0}$ and $P_{\mu^4}$, since we
need to know the angular diameter distance and $H(z)$ to be able to
convert $\uvec$ to $\kvec$.

If the cosmology assumed in the conversion from $\uvec$ to $\kvec$ is
incorrect, there will be an asymmetry in the measured depth versus
measured angular size of an object \citep{alcock79}.  As a result,
features, such as the effective bubble size, will appear distorted in
the 21 cm map [the AP effect].  Tests for this asymmetry can be used
to put constraints on cosmological parameters.  This effect is
implicitly included in the analysis in \S \ref{mu4} because we vary
the cosmological parameters in the conversion from $\uvec$ to $\kvec$.
To extract cosmology from $P_{\mu^4}$ alone, it is more subtle to
include this effect.

The presence of this new angular dependence from the AP effect will
contribute an additional term to the $\mu$ decomposition of $P_{\Delta
T}$ (equation \ref{pseqn}), which has a unique $\mu^6$ angular
dependence with coefficient \citep{ nusser04, barkanaAP}
\begin{equation}
P_{\mu^6}(\bfk) = -\alpha \, \bar{x}_H^2 \,(4 \, P_{\delta_L \delta_L} - \frac{\partial P_{\delta_L \delta_L}}{\partial \log k}),
\label{eq:mu6}
\end{equation}
keeping terms to linear order in $\alpha$.  Here $1 +\alpha$ is the
ratio of the true value of the Hubble constant times the angular
diameter distance, $H(z) \, D_A(z)$, to the value assumed in the
conversion from $\bfu$ to $\bfk$.  This term arises from the coupling
of the AP effect to $P_{\mu^4}$.  If detected, the $\mu^6$ term would
indicate a clear problem with the assumed cosmological model.  The
$P_{\mu^0}$, $P_{\mu^2}$ and $P_{\mu^4}$ terms are also affected by
the AP effect.  Since we currently cannot model $P_{\mu^0}$ and
$P_{\mu^2}$ when the bubbles are important, the AP effect will not be
detectable from these terms.  However, we can model $P_{\mu^4}$ to
zeroth order in $\alpha$.  If the assumed cosmology is incorrect such
that $\alpha \neq 0$, this term becomes \citep{barkanaAP}
\begin{eqnarray}
P_{\mu^4} &=& P^{\rm tr}_{\mu^4} + \alpha \, \left( 5 \, P^{\rm tr}_{\mu^4} - 2
\, P_{\mu^2} + \frac{\partial P_{\mu^2}}{\partial \log k}
\right) \nonumber \\
&- &\alpha_{\perp} \left(3 \, P^{\rm tr}_{\mu^4} + \frac{\partial \,
P^{\rm tr}_{\mu^4}}{\partial \log k} \right),
 \label{eq:mu4AP}
\end{eqnarray}
where $P^{\rm tr}_{\mu^4}$ is the true $\mu^4$ term (what we measure if
$\alpha = 0$) and $(1 + \alpha_{\perp})$ is the ratio of the assumed
angular diameter distance $D_A(z)$ to its true value.  Since the
$P_{\mu^2}$ is generally larger than $P^{\rm tr}_{\mu^4}$, the deviation from $P^{\rm tr}_{\mu^4}$ can be quite significant.

\begin{figure}
\epsfig{file=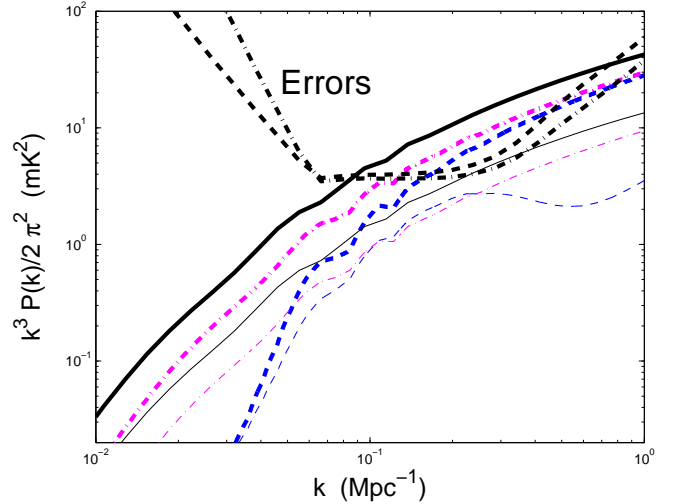, width=8.7cm} 
\caption{Plot of $k^3 P(k)/2 \pi^2$ for $P(k)$ equal to $P^{\rm
tr}_{\mu^4}$ ({\it solid curves}), $P_{\mu^6}$ ({\it dot-dashed
curves}) and $P_{\mu^4} - P^{\rm tr}_{\mu^4}$ ({\it dashed curves})
for $\bar{x}_i = 0.2$ ({\it thick curves}) and $\bar{x}_i = 0.55$ ({\it thin
curves}) at $z = 8$.  Here we assume that $\alpha = 0.1$ and
$\alpha_{\perp} = 0.1$.  The dashed and dot-dashed curves, labeled
``Errors,'' are SKA's errors for a $2$ yr observation on $P_{\mu^4}$
and $P_{\mu^6}$ when $\bar{x}_i = 0.2$, computed with bins $\Delta k =
0.5 \, k$. } \label{fig:AP}
\end{figure}

In Figure \ref{fig:AP}, we plot $P^{\rm tr}_{\mu^4}$ ({\it solid
curves}), $P_{\mu^6}$ ({\it dot-dashed curves}), and $P_{\mu^4} -
P^{\rm tr}_{\mu^4}$ ({\it dashed curves}) for $\bar{x}_i = 0.2$ ({\it
thick curves}) and $\bar{x}_i = 0.55$ ({\it thin curves}) using the
FZH04 analytic model at $z = 8$.  We take $\alpha = 0.1$ and
$\alpha_{\perp} = 0.1$ to roughly match the current uncertainty in
these parameters.  The first generation of 21 cm arrays will not be
sensitive to the AP effect or to $P_{\mu^4}$.  The dashed and
dot-dashed curves in Figure \ref{fig:AP} that are labeled ``Errors''
represent the sensitivity curves for SKA to the $P_{\mu^4}$ and
$P_{\mu^6}$ terms assuming $2$ yr of observation.  SKA will not have
the sensitivity to give a meaningful measurement of either $\alpha$ or
$P_{\mu^4}$, but, after a sufficiently long integration, may be able
to detect $P_{\mu^6}$.  A similar conclusion holds for MWA5000.

Let us now make this analysis more quantitative. If we take the Fisher matrix
$F_{ij}$ at a given $k$ for the parameters $P_{\mu^0}, P_{\mu^2},
P_{\mu^4},$ and $P_{\mu^6}$ (indexed 1 through 4), then marginalize
the contaminated parameters $P_{\mu^0}$ and $P_{\mu^2}$, the Fisher
matrix of just the parameters $P_{\mu^4}$ and $P_{\mu^6}$ is
\begin{equation}
 \bfF'_k = \left( \begin{array}{cc}
{\bfF^{-1}}_{33} & {\bfF^{-1}}_{34}\\
 {\bfF^{-1}}_{43} & {\bfF^{-1}}_{44} \end{array} \right)_k^{-1}.
\end{equation}
By the chain rule, it follows that the Fisher matrix for the cosmological parameters $\lambda_1, ..., \lambda_n$
obtained from just $P_{\mu^4}$ and $P_{\mu^6}$ (indexed by 1 and 2) is
\begin{eqnarray}
F''_{ij}(k) &= & F'_{11} \frac{\partial P_{\mu^4}}{\partial \lambda_i}\frac{\partial P_{\mu^4}}{\partial \lambda_j} + F'_{12} \frac{\partial P_{\mu^4}}{\partial \lambda_i}\frac{\partial P_{\mu^6}}{\partial \lambda_j} \nonumber \\
& + &F'_{21} \frac{\partial P_{\mu^6}}{\partial \lambda_i}\frac{\partial P_{\mu^4}}{\partial \lambda_j} + F'_{22} \frac{\partial P_{\mu^6}}{\partial \lambda_i}\frac{\partial P_{\mu^6}}{\partial \lambda_j}.
\end{eqnarray}

  In Table \ref{table5} we consider the scenario in which $\bar{x}_H =
0.8$ and only $P_{\mu^4}$ and $P_{\mu^6}$ yield a pristine measure of
the linear density power spectrum via equations (\ref{eq:mu4AP}) and
(\ref{eq:mu6}). (While $P_{\mu^4}$ depends on $P_{\mu^2}$ to linear order in $\alpha$, since we can
measure $P_{\mu^2}$, we can always use $P_{\mu^4}$ and $P_{\mu^6}$
to measure $P_{\delta \delta}$.)  We assume that $\alpha$ and
$\alpha_{\perp}$ are zero, but nonzero values do not change the
results significantly.  In this case, MWA5000 and SKA can improve
constraints moderately on cosmological parameters obtained from
current CMB data sets.  However, they are unable to compete with
Planck.  The measurement of $P_{\mu^4}$ and $P_{\mu^6}$, when combined
with Planck, will be most useful for measuring $\bar{x}_H$ rather than
for constraining cosmological models.  Two years of observation with
SKA can constrain $\bar{x}_H$ to better than $10\%$.  Even MWA50K
is not able to sensitively constrain the signal on its own with this
method.

SKA does fare better than MWA5000 for the analysis in this section,
which was not the case in \S \ref{mu4}.  This stems from MWA5000
being significantly more concentrated than SKA and therefore not as
sensitive to small scale modes perpendicular to the LOS.  These
modes are important to be able to separate the different $\mu$ terms.
Our design for MWA50K is very concentrated, like MWA5000, and so is
also not optimal for measuring $P_{\mu^4}$ and $P_{\mu^6}$.  

If one assumes some simple parameterization for $P_{xx}$ and $P_{x \, \delta}$
(or $P_{\mu^0}$ and $P_{\mu^2}$), rather than marginalizing over
$P_{\mu^0}$ and $P_{\mu^2}$ for each $k$-bin as is done here, then perhaps
one could measure $P_{\delta \delta}$ with more confidence.

\begin{deluxetable*}{l c c c c c c c c c}
\tabletypesize{\footnotesize}
\tablecaption{Errors on cosmological parameters estimates from 21 cm
observations when only
$P_{\mu^4}$ and $P_{\mu^6}$ are not contaminated by the
bubbles.\tablenotemark{a}}
\tablehead{ & \colhead{$\tau$} & \colhead{$\Omega_\Lambda$} & \colhead{$\Omega_m \,h^2$} & \colhead{$\Omega_b
 \,h^2$} & \colhead{$n_s$} & \colhead{$\delta_H \times 10^5$} & \colhead{$\alpha_s$} & \colhead{$\Omega_\nu$} &
\colhead{$\bar{x}_H(z = 8)$}}

\startdata
  & 0.1 & 0.7 & 0.14 & 0.022 & 1.0 & 3.9 & 0.0
 & 0.0 & 0.8\\ \hline
CCMB & 0.060 & 0.084 & 0.017 & 0.0014 & 0.07 & 0.3 & 0.039 & 0.12 & -\\
CCMB + MWA5000 & 0.058 & 0.066 & 0.011 & 0.0012 & 0.05 & 0.2 & 0.033 & 0.05 & 0.3\\
CCMB + SKA & 0.057 & 0.062 & 0.011 & 0.0012 & 0.04 & 0.2 & 0.028 & 0.04 & 0.3\\
\hline Planck &
 0.005 & 0.011 & 0.0023 & 0.00017 & 0.0047 & 0.03 & 0.007 & 0.010 & -
 \\
Planck + MWA5000 & 0.005 & 0.011 & 0.0022 & 0.00017 & 0.0047 & 0.03 & 0.007  & 0.010 & 0.07 \\
Planck + SKA & 0.005 & 0.011 & 0.0022 & 0.00017 & 0.0047 & 0.03 & 0.007  & 0.010 & 0.07 \\
Planck + MWA50K & 0.005 & 0.010 & 0.0020 & 0.00017 & 0.0047 & 0.03 & 0.007  & 0.009 & 0.06\\ 
\enddata

\label{table5}
\tablenotetext{a}{See table note (a) in Table \ref{table3} for the specifications used in these calculations.  Observations with SKA and MWA50K are for 400 hours on ten places in the sky.}
\end{deluxetable*}

\subsection{Large Scales}
\label{largescales}
Up until now, we have ignored all components of the signal that are
contaminated by the bubbles.  On large scales, this may not be
necessary.  On scales much larger than
the effective HII bubble size $R_{\rm eff}$, if the Poisson fluctuations owing to the bubbles are unimportant, the bubble fluctuations will trace the density fluctuations.  Thus, when $k \ll R_{\rm eff}^{-1}$, we may have the relations
\begin{equation}
P_{xx}(k) \approx b_1^2 \, P_{\delta \delta}, ~~~ ~~~P_{x \delta}(k) \approx b_1
\, P_{\delta \delta}.
\label{eq:lslimit}
\end{equation}
It is not necessarily the case with HII bubbles during reionization,
as it is typically for galaxy surveys, that Poisson fluctuations are
unimportant at large scales.  If we include both the part of the
signal that traces the $P_{\delta \delta}$ and the Poisson
component, we can write the 21 cm power spectrum at large scales as
\begin{eqnarray}
P_{\Delta T}({\bfk}) & =  & \tilde{T}_b^2 \, \{ \left[\bar{x}_H^2
 +  b_1^2 - 2 \bar{x}_H \, b_1 \right] + 2 \mu^2 \,\nonumber \\
& &\times  \left[ \bar{x}_H^2 -  \bar{x}_H \, b_1 \right] +
\mu^4 \,\left[ \bar{x}_H^2  \right] \} P_{\delta \delta}  + P_{\rm poi},
\label{pslseqn}
\end{eqnarray}
where $P_{\rm poi}$ is the Poisson contribution, which is constant in
$k$.  This type of decomposition may also hold when spin temperature
fluctuations are important.  On large scales we may
again parameterize spin temperature fluctuations with the relation
$\delta_{T_s}= b_{T_s} \, \delta$ for some constant $b_{T_s}$, and
again the 21 cm power spectrum is proportional to $P_{\delta \delta}$.
If we include the AP effect, terms enter with derivatives of
$P_{\delta \delta}$.

Equation (\ref{pslseqn}) is promising in that, when Poisson fluctuations are unimportant, there are effectively three unknowns
($b_1, \bar{x}_H$ and $P_{\delta \delta}$) and three equations, since
the $\mu^0, \mu^2$ and $\mu^4$ components can, in principle, be
measured.  All three of the unknowns are very informative:
$\bar{x}_H(z)$ tells us about the global reionization history,
$b_1(z)$ indicates where the bubbles are located (i.e.  overdense or
average density regions), and of course $P_{\delta \delta}$ is quite
interesting.  In addition, $b_1$ can be much larger than unity, such that the
signal is enhanced over that of a neutral medium.

There are several difficulties with extracting cosmology from these
large scale modes. One difficulty is that modes larger than the width
of the 21 cm survey will be contaminated by foregrounds.  Another
complication is that 21 cm interferometers will have progressively
more trouble capturing terms of increasing order in $\mu$, which is
necessary to separate the terms in equation (\ref{pslseqn}).  Despite
these difficulties, it is probable that modes on scales near the
baryonic wiggles ($k \sim 0.1 ~ \Mpc^{-1}$) will be in this large
scale regime.  This is a region in k-space that contains much
cosmological information and is also on scales where interferometers
will be most sensitive to the $\mu$ decomposition of the signal
(Fig. \ref{fig:mudecomp}).

\section{Discussion}
\label{Discussion}

In this paper, we have used the FZH04 analytical model of reionization
to calculate the power spectrum of 21 cm brightness temperature
fluctuations $P_{\Delta T}(\bfk)$, extending this calculation beyond
calculations done in FZH04 by including redshift-space
distortions. When $\bar{x}_i \la 0.5$ or on smaller scales than the
effective bubble size, these distortions are quite important and give
the power spectrum a substantial anisotropy between modes parallel and
perpendicular to the LOS. These distortions not only increase the
signal, but may allow us to separate $P_{xx}$, $P_{x\delta}$ and
$P_{\delta \delta}$, facilitating the measurement of the size and bias
of the HII bubbles and perhaps the spectrum of density fluctuations in
the Universe.  We show that higher order terms may complicate the
separation of $P_{\Delta T}$ when the ionized fraction is significant.

To quantify the detectability of $P_{\Delta T}(\bfk)$, and in
particular $P_{\delta \delta}$, we make realistic sensitivity
estimates for LOFAR, MWA and SKA.  The most important parameter for
these interferometers is the collecting area.  But, this is not to say
that the other factors that go into the design are unimportant.  We agree
with the conclusion of \citet{bowman05} that, everything else being
equal, arrays with denser cores will be more sensitive to $P_{\Delta
T}(\bfk)$.  This is because modes along the LOS can be detected by
even the shortest baselines, and arrays with cores have more of these
shorter baselines.  The antenna size can also have a similar effect:
arrays that have large antennae cannot pack them as closely together
as arrays with small antennae.  As a result, they will not sample the
shorter baselines as well.  Smaller antennae also provide a larger
FOV, which will aid statistical detections of the signal.  Because
the current design for LOFAR does not include the shorter baselines
that the design for MWA has and because the design for LOFAR results
in a much smaller FOV, we find that, despite
differences in collecting area, LOFAR and MWA will be comparably
sensitive to $P_{\Delta T}(\bfk)$ at most redshifts. This is not to
say that this will be the case when these instruments are
actually deployed.  Since none of the discussed 21 cm arrays have begun construction, their designs can still be optimized.

Even with an optimally constructed radio interferometer, the removal
of foregrounds that are $10^4$ times larger than the 21 cm
fluctuations will be a serious challenge.  In this paper, we find that
foregrounds will contaminate the signal on scales greater than the
depth of the slice used to construct the 21 cm power spectrum.  At
most scales smaller than this, we are optimistic that foregrounds can
be cleaned below the signal.  On such scales, we show that fitting a
quadratic or cubic polynomial to the observed visibilities in the frequency
direction has little difficulty cleaning a realistic model for
foregrounds.  It does not appear to be the case, as was claimed by
\citet{oh03} and \citet{gnedin04}, that foregrounds will contaminate
all angular modes beyond repair.

Applying our calculation for the detector noise and foreground power
spectrum, we find that MWA and LOFAR will not be sensitive to the
$P_{\mu^4}$ component of $P_{\Delta T}(\bfk)$.  This component is
particularly interesting because it traces the linear-theory density
power spectrum.  However, these interferometers will be sensitive to
$P_{\mu^0}$, which probably will tell us more about
the astrophysics of reionization than about cosmology, except perhaps
for very small ionization fractions.  MWA5000 and SKA will be
moderately sensitive to $P_{\mu^4}$ and $P_{\mu^6}$, but not sensitive
enough to provide competitive constraints on cosmological parameters.
We find that only if there exists a time when density fluctuations
dominate $P_{\Delta T}$, will upcoming probes of 21 cm emission be
able to place competitive constraints on cosmological parameters.  In
addition, planned 21 cm interferometers will not be very sensitive to
the signal for $z \ga 12$.  The is primarily because detector noise
fluctuations are proportional to $T_{\rm sky}(z)$, which scales as $(1
+ z)^{2.6}$.

If there is a period where density fluctuations dominate $P_{\Delta
T}$, a $2$ yr observation with MWA5000 plus Planck can give the
constraints $\delta \Omega_{w} = 0.0017$ (a $1.7$ times smaller
uncertainty than from Planck alone), $\delta w = 0.06$ ($1.5$ times),
$\delta \Omega_m h^2 = 0.0009$ ($2.5$ times), $\delta \Omega_b h^2 =
0.00012$ ($1.5$ times), $\delta n_s = 0.0033$ ($1.4$ times), $\delta
\alpha_s = 0.003$ ($2.7$ times), $\delta \Omega_{\nu} = 0.003$
($3$ times) and $\delta x_H = 0.03$. SKA plus
Planck yield similar constraints and MWA50K can do even better.
However, if $\tau = 0.17$, as suggested by WMAP, and reionization began
at $z \approx 20$, observations of the signal at scales much larger
than the effective bubble size may be the most promising direct method
to probe cosmology (\S \ref{largescales}).

Observations must overcome many additional challenges beyond those
that have been discussed in this paper.  Issues that we have not
addressed include contamination by radio recombination lines,
terrestrial radio interference, the residuals left from wave front
corrections for a turbulent atmosphere, and the enormous data
analysis pipeline needed to analyze potentially larger data sets
than those from current experiments.  The 21 cm signal will
also be affected by gravitational lensing by intervening material
\citep{zahn05, mandel05}.  If taken into account, this effect can
further improve cosmological parameter estimates \citep{zahn05}.

Cosmic variance sets a limit on how well we can constrain cosmological
models with the CMB.  Because 21 cm emission can be observed as a
function of redshift, this signal allows us to measure many more
independent modes than is possible with the CMB.  We have seen that
cosmological parameters are extractable from 21 cm emission.  In an
ideal case in which reionization begins at relatively low redshifts,
upcoming interferometers may be able to compete with future CMB
experiments such as Planck.  If reionization begins at higher
redshifts or if the spin temperature fluctuations are important, a
more sensitive interferometer will be required than those that are
currently planned to be able to compete with CMB parameter
constraints.  Regardless of how reionization actually proceeded,
high-redshift 21 cm emission has the potential to become a valuable
probe of cosmology.\\ \\

We thank Judd Bowman, Bryan Gaensler, Adam Lidz and Miguel Morales for useful
discussions.  This work was supported in part by NSF grants ACI
96-19019, AST 00-71019, AST 02-06299, and AST 03-07690, and NASA ATP
grants NAG5-12140, NAG5-13292, and NAG5-13381.

\begin{appendix}

\section{The Effect of Evolution}
\label{evolution}

Observations must have a large enough bandwidth to provide adequate
signal-to-noise, but the larger the bandwidth, the more the signal
will evolve along the LOS direction of the 21 cm map.  During
reionization this evolution includes (1) density inhomogeneities
growing with time and (2) the average 21 cm brightness temperature
declining as the Universe expands and the bubbles grow and occupy
progressively more space.  The latter should dominate over the former
owing to the relatively short timescales during which the ionization
fraction changes by order unity. Since the Universe is evolving across
modes along the LOS and is not evolving for those transverse to the
LOS, this evolution will break the spherical symmetry of the signal
and potentially make it more difficult to observe $P_{\mu^0}$,
$P_{\mu^2}$, $P_{\mu^4}$ and $P_{\mu^6}$.  In this section, we attempt
to quantify the potential size of this effect for an idealized case.

Very simple models for reionization set $P_{\Delta T} = b^2 \,
P_{\delta \delta}$ where $b$ is the effective bias.  This
parameterization is certainly not correct, and the bias will have a
scale dependence for scales around the size of the bubbles.  This
model for the power spectrum is reasonable at the beginning of
reionization, when density fluctuations dominate, and on much larger
scales than the bubbles .  Fortunately,
these are the regimes in which the cosmological information is most readily
extractable from the signal (see \S \ref{mu4} and \S \ref{largescales}).

With the assumption that $P_{\Delta T} =b^2 \, P_{\delta \delta}$, we
can write an expression for the power spectrum in a region of
width $2 \,\Delta r$
\begin{equation}
P_{\Delta T}({\bfk}) = 
\langle \frac{1}{\Delta V} \, \mathop{\int_{\Delta V}}  d^3\bfr d^3\boldsymbol{\epsilon} \,  F(r_{||}) \,
\delta_0({\bfr}) \, F(r_{||} + \epsilon_{||})\,\delta_0({\bfr} +
{\bf \epsilon}) \exp(-{\it i} {\bfk}\cdot {\bf \epsilon}) {\rangle},
\label{eqn:evol}
\end{equation}
where $\Delta V$ is specified in the LOS direction by the conditions
${|r_{||} - \bar{r}_{||}| < \Delta r}$ and ${|r_{||} + \epsilon_{{||}}
- \bar{r}_{||}| < \Delta r}$ and $r_{||}$ is the projection of $r$
along the LOS and is assumed to be much larger than $\Delta r$ in the
angular directions..  To constants of order unity, $F = b(z) \,
(1+z)^{-1/2}$ [the factor of $(1+z)^{-1/2}$ owes to the evolution of
$\bar{T}_b$ as well as the growth factor], and $\delta_0$ are the
linear density fluctuations at $z = 0$ [$\delta_L(z) = G(z)
\,\delta_0$, where $G$ is the growth factor]. Here the subscript
${||}$ indicates the LOS direction. The $\langle \rangle$ indicates an
ensemble average of maps, and $\langle \delta_0(\bfr) \delta_0(\bfr +
{\bf \epsilon}) \rangle = \zeta_{\delta_0 \delta_0}({\bf \epsilon})$
in which $\xi_{\delta_0 \delta_0}$ is the linear density field
correlation function.  The Fourier transform of $\xi_{\delta_0
\delta_0}$ is $P_{\delta_0 \delta_0}$.  By the Fourier transform
convolution identity, equation (\ref{eqn:evol}) is equivalent to
\begin{equation}
P_{\Delta T}({\bfk}) = \int dk_{||}'\, \tilde{W}(k_{||}') \, P_{\delta_0
\delta_0}({\bfk} - {\bfk}'),
\label{eq:conv}
\end{equation}
where
\begin{equation}
\tilde{W}(k_{||}) = \frac{1}{2 \, \Delta r}\int_{\Delta V_{||}} dr_{||} \, d\epsilon_{{||}} \,F(r_{||}) \, F(r_{||} + \epsilon_{{||}}) \, \exp(-{\it i} k_{||} \, \epsilon_{{||}}).
\label{eq:wfunc}
\end{equation}

Prior to reionization, $b = 1$ and $\tilde{W}(k)$ is almost identical to the
window function for a top-hat in real space.  Since this window
function peaks when its argument is near zero, for large $\bfk$, then
$P_{\Delta T}({\bfk}) \approx P_{\delta_0 \delta_0}(\bfk) \, \int dk'
\,\tilde{W}(k')$. This is exactly what we
should expect; in this limit, we are effectively averaging over modes
located at different redshifts, such that only the amplitude of the
observed power spectrum is affected by evolution.  The difference
between this window function for a universe in which the ionized
fraction is and is not changing will indicate the degree to which the
spherical symmetry is affected by the evolving ionized fraction.  For
example, if the window function becomes less peaked, then the
spherical symmetry of more modes is affected by the evolving ionized
fraction. In cases in which the window function is unaffected, this means
that evolution only affects the normalization of the power spectrum
through a factor of $\int dk' \,\tilde{W}(k')$.  While the normalization is
also interesting, it is degenerate with the parameters $\bar{x}_H$, $b_1$
(the linear bias of the bubbles) and $\sigma_8$ and therefore is less
important.

\begin{figure}
\begin{center}
\epsfig{file=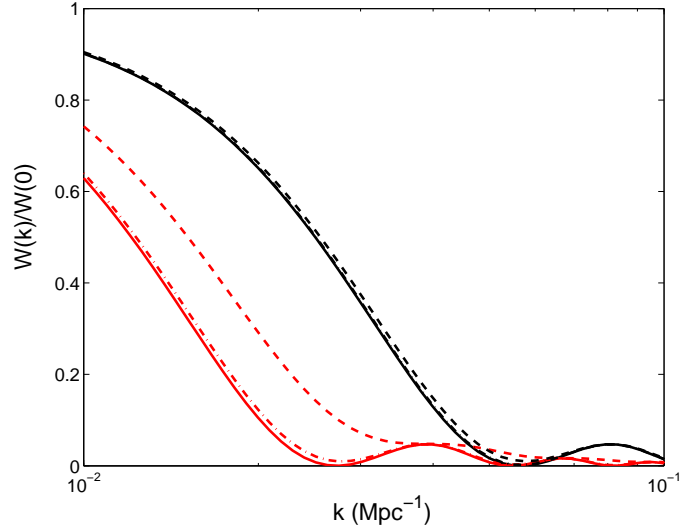, width=9cm}
\end{center}
\caption{Window function $\tilde{W}(k)$, defined in equation
(\ref{eq:wfunc}), for maps of comoving depths $110 ~ \Mpc$ ({\it outer
curves}) and $230$ Mpc ({\it inner curves}) at $z = 10$ .  We plot the
window function for a universe prior to reionization $b(z) = {\rm
constant}$ ({\it solid curves}) and for when the universe is starting
to be ionized (quickly) with $b(z) \propto [m(z -10) + 1]$ for $m =
0.5 \, {\rm ~ and ~} \, 1$ ({\it dot-dashed and dashed curves,
respectively}).}
\label{fig:winfunc}
\end{figure}

Figure \ref{fig:winfunc} plots the window function at $z = 10$ for
maps with comoving depth $110$ and $230 ~ \Mpc$ ($\Delta z = 0.5$ and
1) for a neutral universe ({\it solid curve}) and two cases in which $b(z)
\propto [m(z -10) + 1]$ for $m = 0.5$ and $1$.  This is a conservative
choice; such a quick change of $b$ is much larger than models predict at
the beginning of reionization.  Figure
\ref{fig:winfunc} demonstrates that the window function is virtually
unaltered for three of the four cases.  Only in the most extreme case,
where $m(z -10) + 1$ varies from 0 to 1 within the map, is the
window function significantly altered. Therefore, the evolution of the
signal for this simple model weakly breaks the spherical symmetry of
the signal.

\section{Angular Averaged Sensitivity}
\label{Ap:spav}
 The spherically averaged signal is obtained by summing up all pixels
in a shell with the same $|\bfk|$. If $\Delta k = \epsilon k$, it
follows from equation (\ref{eq:dP}) that the error on $P_{\Delta
T}(k)$ from a measurement of all pixels in a shell with constant $k$
is given by
\begin{eqnarray}
\delta P_{\Delta T}(k)& =&  \left[ \sum_{\theta} ~\left(
\frac{1}{\delta P_{\Delta T}(k, \theta)} \right)^2 \right]^{-1/2}  \nonumber \\
& \approx & \left[ k^3 \, \int_{\arccos[\min(\frac{y k}{2 \pi},
1)]}^{\arcsin[\min(\frac{k_*}{k}, 1)]} ~ d \theta \,\sin(\theta) \, \frac{1
}{\left(D \, P_{\rm 21 cm}(k) + \frac{E}{n(k
\sin(\theta))}\right)^2} \right]^{-1/2}, \label{eq:dPk}\\
& & D = \sqrt{\frac{(2 \pi)^2 \, A_e}{\lambda^2 \, x^2 \, y \, \epsilon}}
~ ~ ~ ~E = \frac{ 2 \pi \, \sqrt{x^2 y} \, \lambda^3 \, T_{\rm
sys}^2}{\sqrt{ \, \epsilon} \, A_e^{3/2} \, (B \,t_0)}, \label{eq:C}
\end{eqnarray}
where $k_*$ corresponds to the longest wavevector perpendicular to the
LOS probed by the array.  In equation
(\ref{eq:dPk}), the lower bound of the integrand reflects the sharp
cutoff in the number of pixels imposed for wavelengths that do not fit
in the width of the box, $y$.  Some interferometers, such as LOFAR, will be
able to observe $N_{\rm point}$ separate fields of view
simultaneously.  To include this effect, one can simply divide
equation (\ref{eq:dPk}) by $\sqrt{N_{\rm point}}$.

To gain intuition into the scalings of equation (\ref{eq:dPk}), let us take a top-hat distribution of baselines
with density $\rho(\lambda)$ in the limit in which the detector
noise dominates such that we can set $P_{\rm 21 cm} = 0$ in equation
(\ref{eq:dPk}).  In this case, we can evaluate the integral in
equation (\ref{eq:dPk}), which yields
\begin{eqnarray}
\delta P_{\Delta T}(k) & =&  E \, \left(\min[\frac{y k}{2 \pi}, 1] \right)^{-1/2} \,\rho^{-1} \, k^{-3/2} ~~~~ k < k_* \\
&=& E \, \rho^{-1} \, k^{-3/2} \,(1 - \frac{\sqrt{k^2 - k_*^2}}{k})^{-1/2} ~~~~ k
> k_*.
\end{eqnarray}
When $k \gg k_*$, $\delta P_{\Delta T}(k) \propto k^{-0.5}$.

\end{appendix}


\end{document}